\documentclass[fleqn,usenatbib]{mnras}

\usepackage{newtxtext,newtxmath}

\usepackage[T1]{fontenc}

\DeclareRobustCommand{\VAN}[3]{#2}
\let\VANthebibliography\thebibliography
\def\thebibliography{\DeclareRobustCommand{\VAN}[3]{##3}\VANthebibliography}

\usepackage{graphicx}	
\usepackage{amsmath}	

\usepackage{caption}
\usepackage{subcaption}

\usepackage{graphicx}
\usepackage{float}
\usepackage{amsmath}
\usepackage{xcolor}
\usepackage{multirow} 
\mathchardef\mhyphen="2D 

\usepackage{tikz} 
\usetikzlibrary{shapes.geometric, arrows} 
\usepackage{adjustbox}
\usepackage{booktabs} 
\tikzstyle{datanode} = [rectangle, rounded corners, 
minimum width=3cm, 
minimum height=1cm,
text centered, 
draw=black, 
text width=4cm, 
fill=red!30]
\tikzstyle{narrowdatanode} = [rectangle, rounded corners, 
minimum width=2cm, 
minimum height=1cm,
text centered, 
draw=black, 
text width=3cm, 
fill=red!30]
\tikzstyle{methodnode} = [rectangle, rounded corners, 
minimum width=3cm, 
minimum height=1cm,
text centered, 
draw=black, 
text width=4cm, 
fill=blue!30]
\tikzstyle{narrowmethodnode} = [rectangle, rounded corners, 
minimum width=2cm, 
minimum height=1cm,
text centered, 
draw=black, 
text width=3cm, 
fill=blue!30]

\usepackage{amsmath}
\usepackage{cuted}
\tikzstyle{arrow} = [thick,->,>=stealth]

\newcommand*{\cosmo}{\ensuremath{\mathbf{\Omega}}}
\newcommand*{\posterior}{\ensuremath{P(\cosmo|\mathbf{D})}}
\newcommand*{\posteriorone}{\ensuremath{P(\cosmo|{D_i})}}
\newcommand*{\prior}{\ensuremath{P(\cosmo)}}
\newcommand*{\zL}{\ensuremath{{z_{\rm{L}}}}}
\newcommand*{\zS}{\ensuremath{{z_{\rm{S}}}}}
\newcommand*{\zi}{\ensuremath{{z_{\it{i}}}}}
\newcommand*{\zLi}{\ensuremath{{z_{\rm{L}\it{,i}}}}}
\newcommand*{\zSi}{\ensuremath{{z_{\rm{S}\it{,i}}}}}
\newcommand*{\zLobs}{\ensuremath{z_{\rm{L,obs}}}}
\newcommand*{\zSobs}{\ensuremath{z_{\rm{S,obs}}}}
\newcommand*{\zLobsi}{\ensuremath{{z_{\rm{L,obs}\it{,i}}}}}
\newcommand*{\zSobsi}{\ensuremath{{z_{\rm{S,obs}\it{,i}}}}}
\newcommand*{\ztruei}{\ensuremath{{z_{\rm{True}\it{,i}}}}}
\newcommand*{\zLtruei}{\ensuremath{{z_{\rm{L,True}\it{,i}}}}}
\newcommand*{\zStruei}{\ensuremath{{z_{\rm{S,True}\it{,i}}}}}
\newcommand*{\ri}{\ensuremath{r_i}}
\newcommand*{\robsi}{\ensuremath{r_{\rm{obs}\it{,i}}}}
\newcommand*{\robs}{\ensuremath{r_{\rm{obs}}}}
\newcommand*{\DL}{\ensuremath{D_{\rm{L}}}}
\newcommand*{\DLS}{\ensuremath{D_{\rm{LS}}}}
\newcommand*{\DS}{\ensuremath{D_{\rm{S}}}}
\newcommand*{\Lens}{\ensuremath{\rm {L}}}
\newcommand*{\Nonlens}{\ensuremath{\rm{\hat{L}}}}
\newcommand*{\PLi}{\ensuremath{{P_{\rm{L}\it{,i}}}}}
\newcommand*{\PL}{\ensuremath{{P_{\rm{L}}}}}

\newcommand*{\gammalens}{\ensuremath{\gamma_{\rm{lens}}}}
\newcommand*{\thetaE}{\ensuremath{\theta_E}}

\newcommand*{\rt}{\textcolor{black}}

\newcommand{\NNone}{\texttt{FID}}
\newcommand{\NNtwo}{\texttt{SEE}}
\newcommand{\NNthree}{\texttt{NNB}}
\newcommand{\NNfour}{\texttt{LLT}}

\newcommand{\lcdm}{\ensuremath{\Lambda\rm{CDM}}}

\title[Lens Modeling and Cosmology with Impure Data]{Lens Modeling and Cosmological Inference from an Impure Sample of Galaxy-Galaxy Strong Lenses}

\author[Holloway et al.]{Philip Holloway,$^{1,2}$\thanks{philip.holloway@port.ac.uk}
Aprajita Verma$^{2}$,
Philip J. Marshall$^{3,4}$,
Padmavathi Venkatraman$^{5}$,
\newauthor
Sydney Erickson$^{3,4}$,
Tian Li$^{1}$,
Simon Birrer$^{6}$,
Steven Dillmann$^{3,4}$,
Thomas E. Collett$^{1}$,
\newauthor
and the LSST Dark Energy Science Collaboration\\
$^{1}$Institute of Cosmology and Gravitation, University of Portsmouth, Burnaby Road, Portsmouth, UK\\
$^{2}$Department of Physics, University of Oxford, Keble Road, Oxford, UK\\
$^{3}$Kavli Institute for Particle Astrophysics and Cosmology, Department of Physics, Stanford University, Stanford, CA 94309, USA\\
$^{4}$SLAC National Accelerator Laboratory, 2575 Sand Hill Road, Menlo Park, CA 94025, USA\\
$^{5}$Department of Astronomy, University of Illinois Urbana-Champagin, Urbana, IL 61801, USA\\
$^{6}$Department of Physics and Astronomy, Stony Brook University, Stony Brook, NY 11794,USA
}

\date{Accepted XXX. Received YYY; in original form ZZZ}

\pubyear{\the\year{}}

\begin{document}
\label{firstpage}
\pagerange{\pageref{firstpage}--\pageref{lastpage}}
\maketitle

\begin{abstract}
The start of the Legacy Survey of Space and Time marks a new era for strong lensing science, where the number of strong lenses identified is expected to increase to $\mathcal{O}(10^5)$. In this paper we use a neural network to determine the precision with which lens parameters can be determined, using realistic simulated LSST lensed systems. We find that the Einstein radius can be measured with a mean precision of $3.7\%$ with calibrated uncertainties accurately reflecting the corresponding measurement error.
Based on the performance of current strong lens classifiers, the $\sim 100,000$ detectable strong lenses are expected to be accompanied by a similar or larger number of false positives (non-lenses). In readiness for this we introduce a formalism, termed `COSMIC-BEAMS', to infer cosmological parameters while accounting for contamination by false positives. As a proof-of-concept, using simulated LSST measurements of the Einstein radii of a realistic and impure sample of photometric lens systems, i.e. those without spectroscopic confirmation, we find that the cosmological parameters $\Omega_m$, $\Omega_\Lambda$, and $w$ can be measured to a precision of $0.1$, $0.03$ and $0.15$ respectively for a $w$CDM cosmology. We demonstrate that unbiased cosmological parameters can be inferred even in strong lens samples contaminated by $50\%$ false positives, and that the photometric dataset of $100\,000$ strong lenses will provide equivalent $w$-precision to $2500-3500$ spectroscopic systems. 
\end{abstract}


\begin{keywords}
gravitational lensing: strong  -- cosmological parameters -- methods: statistical
\end{keywords}

\section{Introduction}
Gravitational lensing is the deflection of light emanating from a background source, due to the gravity of a foreground object. Strong gravitational lensing occurs when the deflection is sufficiently large to produce multiple images of the background object, from the perspective of the observer. The background source may be time-varying, such as a quasar, or a `static' galaxy. The deflector may be a galaxy, galaxy group or cluster of galaxies. Strong lens systems are rare (roughly 1 in 2000 galaxies with $I_{\rm{E}}<22.5$ in \emph{Euclid}, \citealp{Walmsley2025}) due to the close angular alignment required between the lens and source galaxies to cause detectable deflection.

Strong lens systems are useful probes of cosmology and galaxy evolution. Time-varying sources such as lensed quasars or supernovae can provide constraints on the Hubble Constant $H_0$ (e.g., \citealp{Wong2020_H0LiCOW_XIII, Shajib2023_TDCOSMO_XII}) as photons emitted simultaneously but travelling different paths through the lens potential arrive at the observer at different times, a delay which is inversely proportional to $H_0$. With regard to galaxy evolution, samples of galaxy-scale lenses have been used to probe properties such as dark matter substructure \citep{WagnerCarena2023,Powell2023}, the slope of the galaxy mass profile (e.g., \citealp{Shajib2021,Etherington2022_No_lens_left_behind,Etherington2023_Beyond_the_bulgehalo_conspiracy}) the stellar initial mass function (IMF, \citealp{Sonnenfeld2015_SL2S_V}) and the velocity dispersion function (VDF, e.g., \citealp{Davis2003,Chae2010,Geng2021}).

The current standard cosmological model, known as \lcdm, consists of a cosmological constant $\Lambda$ associated with dark energy, cold dark matter (CDM), and ordinary matter. This model shows excellent agreement with many cosmological measurements, such as anisotropies in the Cosmic Microwave Background (CMB, \citealp{Planck2020_VI}), measurements of 3x2pt correlations \citep{Abbott2022_DES_Yr3_Cosmo} and Baryon Acoustic Oscillations \citep{Alam2021_eBOSS}. However, results from the Dark Energy Spectroscopic Instrument (DESI, \citealp{DESI2024,DESI_I_2025,DESI_II_2025}) suggest tension with this standard model when combined with other cosmological probes such as the CMB and so accurate tests of this model are of significant interest.
Having independent probes to measure cosmological parameters 
is also important to identify tensions between measurements. For example, measurements of the Hubble constant (see \citealp{Kamionkowski2023} for a review) and $S_8$ parameter ($= \sigma_8\sqrt{\Omega_m/0.3}$, a measure of the degree of inhomogeneity in the universe
e.g., \citealp{Hildebrandt2017,Planck2020_VI,Joudaki2020}), differ between those from the early universe (i.e., the CMB) and those at low redshifts (e.g., weak lensing in the case of $\sigma_8$ or Type 1a supernovae for $H_0$). Static strong lens systems can also be used as such a cosmological probe. The angular diameter distance $D_A = D_A(z,\Omega)$ is dependent on cosmological parameters $\Omega$. The Einstein radius, $\theta_E$, of a strong lens with enclosed mass $M(\theta)$ is given by:
\begin{equation}
    \theta_E = \sqrt{\frac{4GM(\theta_E)\DLS}{\DS\DL}}
\end{equation}
where $\DLS$, $\DL$ and $\DS$ are the (cosmology-dependent) angular diameter distances between the lens-source, observer-lens and observer-source, respectively. Earlier works such as \citet{Marshall2005} and \citet{Grillo2008} have used this relation to constrain cosmological parameters, though sample size limitations restricted the assumed cosmology to a flat universe or relatively weak constraints. The optical bands of the Legacy Survey of Space and Time (LSST, \citealp{Ivezic2019}) undertaken by the NSF/DOE Vera C. Rubin Observatory, and ESA Euclid Wide Survey are expected to each identify $\mathcal{O}(10^5)$ strong lens systems \citep{Collett2015,Holloway2023_NIR_Rates,Ferrami2024}. A small fraction of these ($\sim10,000$ systems), are expected to be spectroscopically confirmed by the 4MOST Strong Lensing Spectroscopic Legacy Survey \citep[4SLSLS,][]{Collett2023} which will provide lens and source redshifts, as well as velocity dispersion measurements for these lenses, as mentioned above. \citet{Li2024} determined the precision with which the cosmological parameters $\Omega_m$, $\Omega_k$, $\Omega_\Lambda$ and $w$ can be determined from a spectroscopic sample of 10,000 lenses expected from the Euclid Wide Survey and 4SLSLS, i.e., with negligible redshift uncertainty and precise velocity dispersion measurements, $\Delta \sigma_v=10\rm{\,km\,s^{-1}}$. 
They found that $w$ should be determined to greater precision than any other single-probe measurement. 
In this work, we extend the analysis of \citet{Li2024} to include the photometric sample of strong lenses, a much larger sample of systems without spectroscopic measurements or confirmation to help provide even tighter cosmological constraints. This work focuses on single-plane galaxy-galaxy lenses; this is a complementary approach to that taken by \citet{Sharma2026}, who investigated the cosmological constraining power of pairs of galaxy-galaxy lenses with similar photometric deflector properties, forming pseudo double-source plane lenses (PDSPL's). 

The large number of strong lens discoveries anticipated in the coming years will require fast lens modeling to provide measurements of lens properties at scale. Given the rapid evaluation time once trained, machine learning methods such as neural networks are a natural fit for such a challenge and have been tested at scale and with high model complexity \citep{Hezaveh2017,Pearson2019,Pearson2021,Schuldt2021,Poh2022,Schuldt2023_HS_IX,Gentile2023,Erickson2024,Poh2025,Venkatraman2025}. One method for doing this is Neural Posterior Estimation (NPE, \citealp{Lueckmann2017,Papamakarios2018}); training a network to predict the posterior (often with a fixed, e.g., Gaussian, functional form) for the parameters of interest. This has been used with a range of science cases in mind, including on the subhalo mass function \citep{WagnerCarena2023} and time delay cosmography (e.g., \citealp{Erickson2024,Venkatraman2025}). Having a full posterior distribution, rather than a point estimate allows population-level inference, and is used in this work to determine lens parameters.

This work is split into two parts. The first focuses on strong lens modeling via Neural Posterior Estimation on simulated LSST data. The second part uses these results to infer cosmological parameters from a realistic sample of strong lenses comprising those which are spectroscopically-confirmed (hereafter the spectroscopic dataset) and lens candidates that have only photometric data (hereafter the photometric dataset).

\begin {figure*}
\centering
\begin{adjustbox}{width=\textwidth}
\begin{tikzpicture}[node distance=2cm, font=\large]
\node (start) [narrowmethodnode] {\texttt{LensPop}\\(D: Section \ref{S:Data})};
\node (Training) [datanode, right of=start, yshift=2cm, xshift=2cm] {Training Dataset (D: Section \ref{S:TrainingSet_Generation})};
\node (Test) [datanode, right of=start,yshift=-2cm,xshift=2cm] {Lens Test Set (R: Section \ref{S:TP_modelling})};
\node (FPDataGeneration) [narrowdatanode,right of=start, xshift=2cm] {Non-Lens Dataset (R: Section \ref{S:FP_modelling})};
\node (PaltasModelling) [narrowmethodnode,right of=Training, yshift=-2cm, xshift=2cm] {\texttt{paltas} Modelling};
\node (InferenceTestSet) [narrowdatanode,right of=PaltasModelling, xshift=2cm] {Inference Data Vectors\\ (M: Section 
\ref{S:Inference_Dataset_Generation})};
\node (CosmoInference) [narrowmethodnode,right of=InferenceTestSet,xshift=2cm] {\texttt{COSMIC-BEAMS}\\(M: Section \ref{S:COSMICBEAMS_Formulation}, \\R: Section \ref{S:CosmoInference})};
\draw [arrow] (start) |- (Training);
\draw [arrow] (start) |- (Test);
\draw [arrow] (Training) -| (PaltasModelling);
\draw [arrow] (Test) -| (PaltasModelling);
\draw [arrow] (FPDataGeneration) -- (PaltasModelling);
\draw [arrow] (PaltasModelling) -- (InferenceTestSet);
\draw [arrow] (InferenceTestSet) -- (CosmoInference);
\end{tikzpicture}
\end{adjustbox}
\caption{Flowchart of the data generation and analysis described in this work. We label data, method and result sections as \emph{D}, \emph{M} and \emph{R} respectively. Datasets are highlighted in red and data manipulation is shown in blue. \texttt{LensPop} is used to form the lens test set population, and is also used to inform the training data. The latter is used to train \texttt{paltas} networks to model LSST-like lensed images. The trained networks are also applied to images which don't contain a lens to measure their behavior when faced with false positives. These results are used to inform the data vectors generated for the cosmological inference.}
\label{PaperStructure}
\end{figure*}
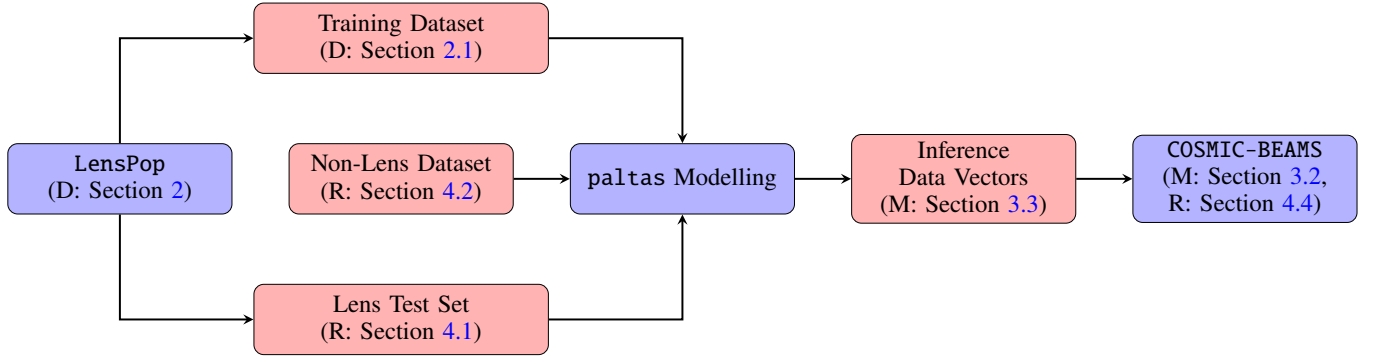

\noindent\textbf{Part 1:} We aim:
\begin{itemize}
    \item \textbf{To determine the precision and accuracy with which lens parameters can be measured from LSST data}, in particular focusing on the Einstein radius $\theta_E$ and mass density slope $\gamma$. We will also determine for which systems these properties can be measured the most accurately.
    \item \textbf{To evaluate the behavior of the neural network when faced with realistic false positives}, i.e., images which do not contain a lens system but which previously confused a lens classifier. By doing this, we will determine whether ML lens modeling could be useful in lens classification, for example in distinguishing whether the predicted lens parameters are within the range expected for real lenses or false positives.
\end{itemize}
\textbf{Part 2:}
The photometric sample of strong lenses candidates in the forthcoming LSST and Euclid Wide Survey is expected to be much larger than the spectroscopic sample by roughly $10:1$. In \citet{Holloway2024} it is shown that even with an ensemble of strong lens classifiers for a sample completeness of $\gtrsim40\%$ the number of false positives will outnumber true positives in the LSST survey. Therefore, to utilize the full strong lens sample, one must account for both increased measurement uncertainties and the effect of contamination from false positives. Similar circumstances affect Type 1a supernova cosmology, for which 
the Bayesian Estimation for Multiple Species (BEAMS) framework was developed \citep{Kunz2007,Hlozek2012,Kunz2013}, to account for impurities in the supernova sample while inferring unbiased cosmological parameters. This was then extended by \citet{Roberts2017_zBEAMS} to account for redshift uncertainties. In this work, we adapt this framework to the strong lensing case, including uncertainties on velocity dispersion, redshifts and lens/non-lens classification.  Given the results from Part 1, we:
\begin{itemize}
    \item  \textbf{Develop a method to incorporate unconfirmed strong lens candidates in cosmological analysis}. We term this lensing framework COSMIC-BEAMS (COntaminated Strong lensing Measurements to Infer Cosmology - Bayesian Estimation Applied to Multiple Species).
    \item  \textbf{Determine the precision with which $\Omega_m$, $\Omega_\Lambda$, $w_0$ and $w_a$ can be measured from a sample of 100,000 strong lens candidates and 10,000 spectroscopically confirmed systems}. This uses the lens model parameter uncertainties determined in Part 1, as well as the behavior of the neural network to false positives.
\end{itemize}
The work in this paper is intended as a proof of concept to demonstrate that the photometric sample of lenses can provide useful cosmological information. As part of this we make some simplifying assumptions, which would need to be relaxed prior to application on real data. In particular, we assume an isothermal mass model during the cosmological inference. Furthermore, we use simulated input measurements (redshifts, Einstein radii and velocity dispersions) which are on average unbiased, though have non-negligible associated uncertainties. When conducting cosmological inference for the photometric lens sample, we assume that an unbiased measurement of each lens galaxy’s velocity dispersion could be made from the Fundamental Plane relation, with appropriate uncertainties. Such a process would use measurements of the apparent magnitude and redshift of the lens galaxy to determine an approximate velocity dispersion for each system. During the lens modeling stage, the distribution of the training set parameters is centered on the test set mean and associated typical uncertainties from the literature are assumed. We discuss the effects of loosening these assumptions in Section \ref{S: Effect of Bias in the Input Data}.

The paper is structured as follows. In Section \ref{S:Data} we describe the simulated LSST lens sample used throughout this work along with the generation of cutouts and network training sets. In Section \ref{S:Method} we describe the networks trained for lens modeling (\ref{S:Network_Training}) the formalism for deriving cosmological parameters from impure datasets (\ref{S:COSMICBEAMS_Formulation}) and generation of data vectors to apply this (\ref{S:Inference_Dataset_Generation}). We present results in Section \ref{S:Results}, including modeling performance on simulated LSST lenses (\ref{S:TP_modelling}), the network's behavior when applied to false positives (\ref{S:FP_modelling}) and cosmological inference from an impure dataset (\ref{S:CosmoInference}). We discuss these results and further applications of the techniques used in Section \ref{S:Discussion} and conclude in Section \ref{S:Conclusion}. The data flow in this work is depicted in Figure \ref{PaperStructure}.

\section{Data}\label{S:Data}
We created a realistic catalog of strong lenses detectable in LSST, then generated single-band LSST-like images for subsequent analysis. 

The lens catalog was generated using \texttt{LensPop} \citep{Collett2015}, which used distinct lens and source populations, with a velocity dispersion function based on SDSS \citep{Choi2007} and a singular isothermal ellipsoid lens mass model. The source properties were taken from a simulated LSST catalog described in \citet{Connolly2010}, complete to $i\sim27.5$. \texttt{LensPop} also included stringent cuts on the detectability of each lens including $\rm{SNR}=\sum S/\sqrt{\sum N^2}>20$ and a magnification cut of $\mu>3$. 

When generating the above lens catalogue we configured \texttt{LensPop} to match the DP0.2 LSST simulation \citep{DESC2021,Korytov2019}. DP0.2 is a large-scale end-to-end simulation over $\sim300\deg^2$ of the simulated LSST sky, incorporating the survey cadence, multi-band imaging, image processing and catalog generation expected from 5 years of LSST data. However, DP0.2 did not include any strong lens systems hence we injected simulated strong lenses into cutouts from DP0.2. Given we use 5-year coadds in this work, the noise in the cutouts will be larger than that from the full 10-year survey. The lens modeling precision and subsequent cosmological precision presented here is likely to be more conservative than for the full LSST survey in light of this. 

To generate the lens catalogue we adopted $100\times$ 30s \emph{i}-band exposures, with seeing $0.83\arcsec$, gain of $0.7e/\rm{ADU}$, zeropoint of 31.8 (1 count/exposure-time), with a $\rm{SNR}\geq20$ threshold. Around $20\,000$ systems were produced which passed these cuts, with $\rm{SNR}\geq20$ being the most influential detectability constraint \citep{Collett2015}. We note that expectations of $\sim100\,000$ detectable systems in LSST \citep{Collett2015} are for optimally stacked coadds (i.e., including only good seeing exposures) from 10 years of LSST data in $g,r$ and $i$-bands \citep{Collett2015}; we concentrated on systems detectable in the $i$-band without optimal stacking (i.e., in full-coadds) for this work. 

Images of the lens systems described above 
were then simulated by \texttt{paltas} and injected into DP0.2 coadds using the LSST Science Pipeline within \texttt{SLSim}\footnote{\url{https://github.com/LSST-strong-lensing/slsim}} (Khadka et al. in prep.). \texttt{paltas} uses the \texttt{Lenstronomy} package \citep{BirrerAmara2018,Birrer2021} for image generation. Since the DP0.2 simulation produced 5-year coadds, the image noise and depth (and the corresponding modeling precision achieved in this work) will be conservative for the complete LSST survey. Lenses were injected into random patches of this simulated LSST sky and thus the PSF, pixel-exposure maps and pixel-noise maps varied between coadds. We used cutouts of 60 pixels ($12\arcsec$ on a side), with the LSST pixel scale of $0.2\arcsec$. To ensure lenses weren't injected on top of existing galaxies, cutouts with flux $2\sigma$ above the noise level within the central 26 pixels (accounting for the typical lens size and offsetting)
were not used for injection. This was to ensure that the simulations only included single-plane lenses (rather than DSPL’s or systems with deflectors at different redshifts), which are the focus of the cosmological inference in this work. Multi-plane lens systems would need to be treated differently, both during simulation and inference. The injected lens and source galaxies included Poisson noise calculated using the exposure map of the corresponding DP0.2 cutout. Therefore the noise in each coadd pixel reflected the number of single-exposures that pixel in the coadd was generated from.
\begin{figure*}
\centering
\centering
\includegraphics[width=0.8\textwidth]{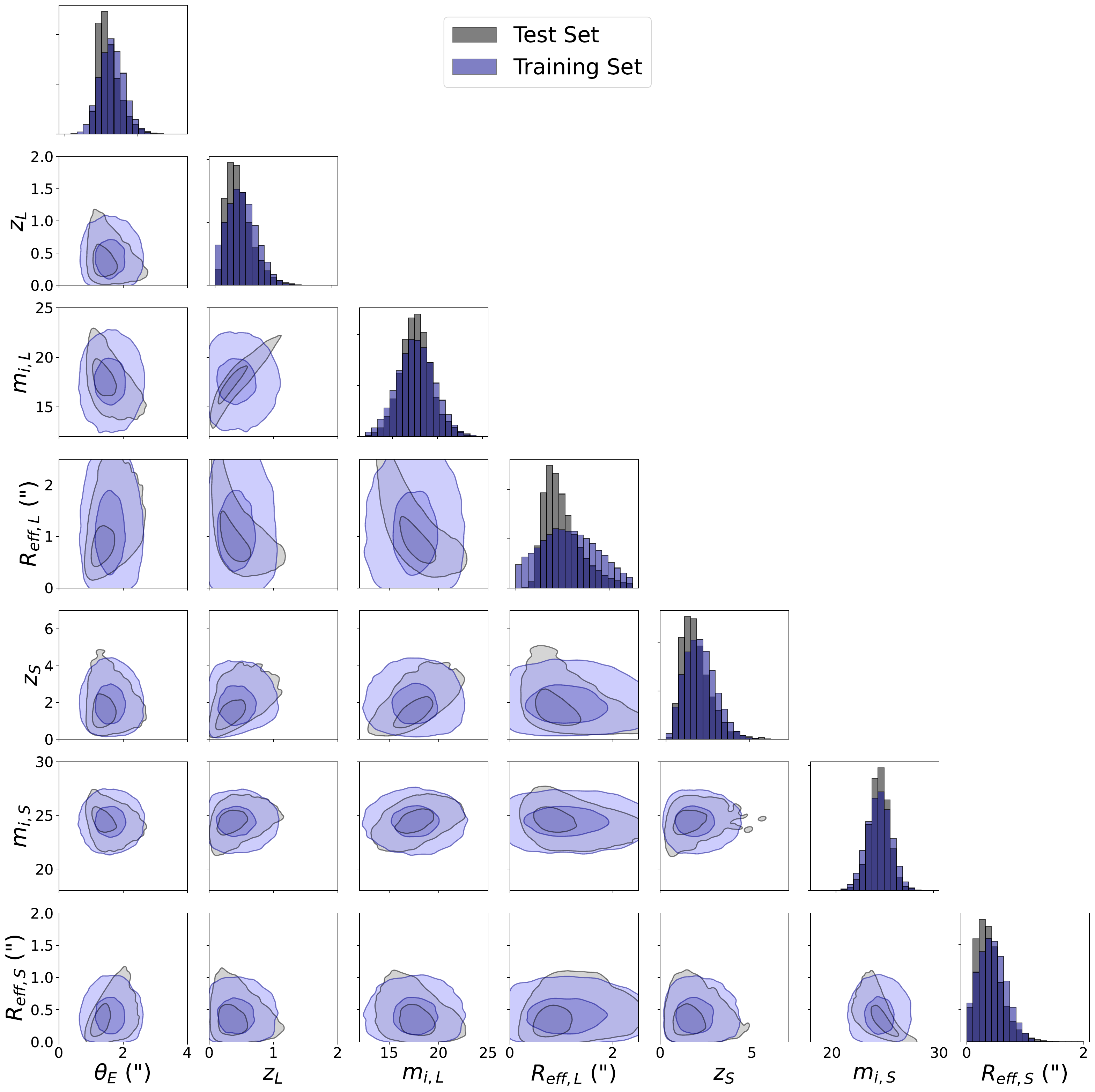}
\caption{Plot of the distributions of lensing parameters of the training set (blue) and test set (grey). With the exception of a magnification cut $\mu>3$, the detection cuts applied to the test set \citet{Collett2015}, were not applied to the training set, which subsequently shows a wider range of lensing systems.}
\label{Train_Test_corner_plot}
\end{figure*}
\subsection{Network Training Set}\label{S:TrainingSet_Generation}
For the training set, we fitted a multivariate Normal (MVN) distribution to key system parameters such as redshifts, Einstein radii, and lens/source magnitudes/sizes of the \texttt{LensPop} test set, and generated a training set by drawing parameter values from a MVN distribution 20\% wider than that of the test set. This ensured the training set encompassed a wide variety of lenses beyond simply those in the test set. The performance of the networks' improved when a magnification cut, calculated from the integrated lensed/unlensed source flux, of $\mu\geq3$ was included in the training data in particular with reduced bias on measuring the Einstein radius, but the training set incorporated none of the co-variances from the test set beyond this and the requirement that $\zL<\zS$ as shown in Figure \ref{Train_Test_corner_plot}. This therefore exposed the networks to a wide range of lens configurations, without fine-tuning them to the test data. 

The lens and source light followed an elliptical S\'ersic profile, while the lens mass distribution in the training set followed a Power-law Elliptical Mass Distribution (PEMD, \citealp{Barkana1998}), with density slope \gammalens{}. For the test set, the power-law index was fixed at $\gammalens{}=2$ (isothermal), as used by \texttt{LensPop}. This ensured that the test set systems were significantly magnified/strongly lensed, while allowing the realistic precision of \gammalens{} measurements to be determined.

\section{Method}\label{S:Method}
\subsection{Neural Network Training}\label{S:Network_Training}
\begin{figure}
 \centering
  \centering
  \includegraphics[width=0.5\textwidth]{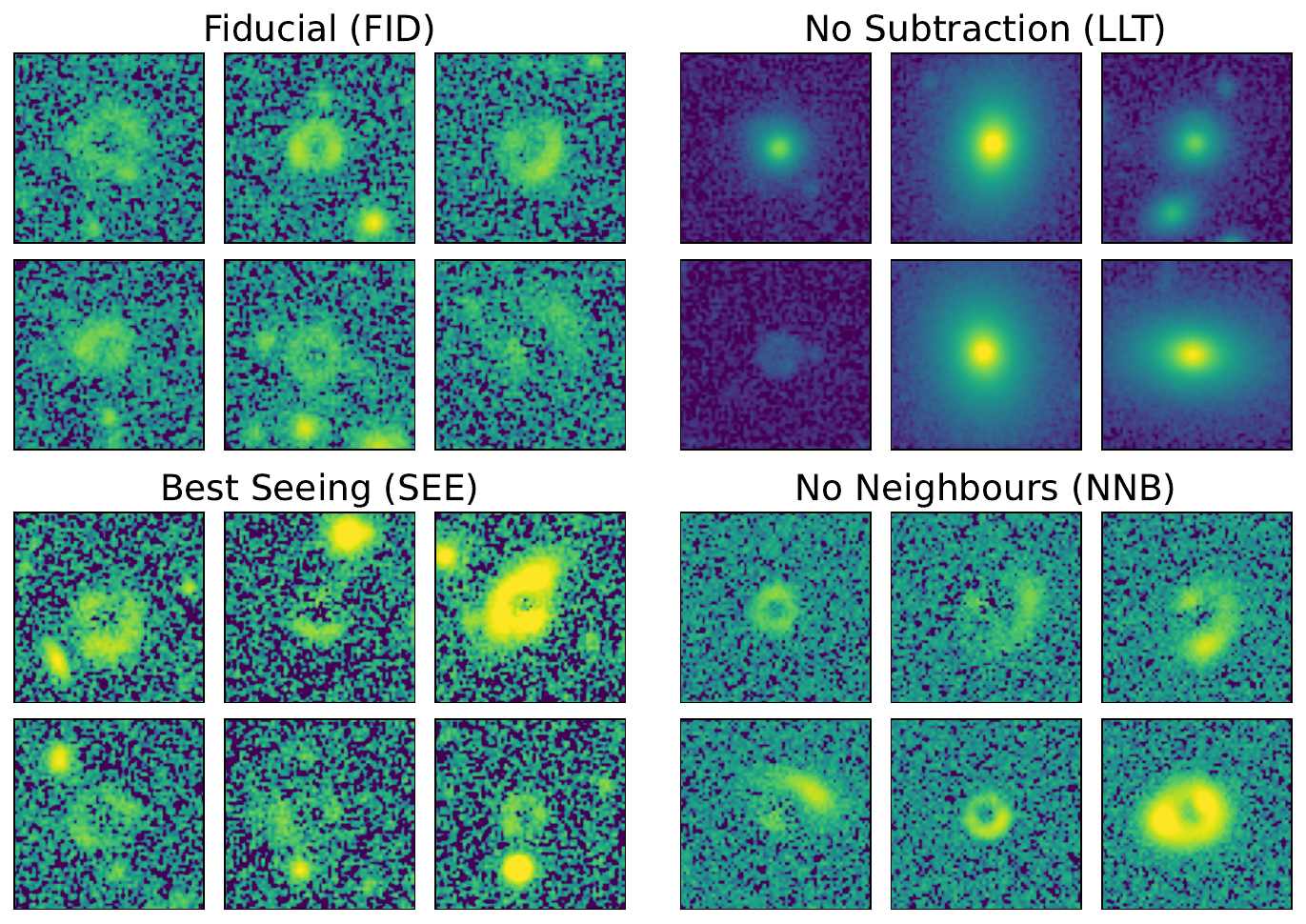}
\caption{Example images used as a test set for each of the 4 networks in this work. Note the \NNthree{} network did not use cutouts from the LSST DP0.2 simulation, and so doesn't contain neighboring (unlensed) objects in the cutout which are visible in the cutouts of the other networks.}
\label{Paltas_collage}
\end{figure}
Four networks were generated using the Neural Posterior Estimation \texttt{paltas} package, each differing by the training set or learning parameters as follows. Except where specified, the networks were trained to infer the mass density slope, shear, lens ellipticity and position, and Einstein radius of the lens. 
    \begin{enumerate}
        \item \textbf{`Fiducial' (\NNone):} Lensed images were injected into the 5-year DP0.2 cutouts. We then subtracted the lens light but retained the Poisson noise from the lens, i.e., we assumed Poisson-limited lens subtraction. This noise was included in both the training and test data. 
        \item \textbf{`Best Seeing' (\NNtwo):} As in the `Fiducial' case, however here the 5-year DP0.2 coadd cutouts consisted only of the top 1/3 best seeing single exposures, rather than all single exposures at a given position. This improved the resolution of the combined image (by $\sim0.1\arcsec$) but reduced the depth.
        \item \textbf{`No Neighbors' (\NNthree):} As in the `Fiducial' case; however, the simulated lensed images were not injected into cutouts from the DP0.2 simulation and consequently did not have neighboring, unlensed, galaxies in the image. Given the DP0.2 cutouts included LSST noise properties, for this network we replicated this by introducing equivalent sky/background noise via \texttt{paltas}. Therefore, these cutouts still had the noise properties expected from 5-year LSST coadds. This network was used to test whether the presence of neighboring objects in the cutout was a significant source of confusion for network modeling.
        \item \textbf{`No Lens Subtraction' (\NNfour)}: As in the `Fiducial' case, but without Poisson-limited lens subtraction. It was this network that was subsequently applied to the false positive systems from the HSC survey (as described later). Additional  parameters (size and magnitudes of the lens and source, and source position) were included in a retrained version of this network in Section \ref{S: Posterior Images}.
    \end{enumerate}
Each network was trained using 500,000 training images, using the Adam Optimizer \citep{Kingma2017} with a learning rate of $5\times10^{-4}$, a batch size of 256 and a diagonal covariance matrix. We show example cutouts from the test sets of each network in Figure \ref{Paltas_collage}.

\subsection{COSMIC-BEAMS Formulation}\label{S:COSMICBEAMS_Formulation}
\begin{figure}
 \centering
  \centering
  \includegraphics[width=0.4\textwidth]{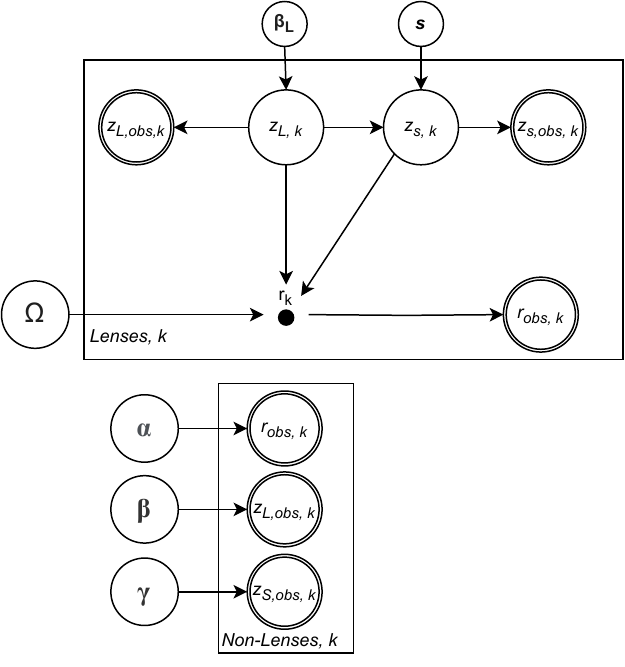}
\caption{Probabilistic Graphical Models (PGM) for the photometric sample formalism, including the effects of contamination. In this diagram, measurements are shown by double circles and inferred or nuisance parameters drawn from a PDF are shown by single circles. However, in the case that this PDF is a delta function (i.e., the $r$ parameter, which for a lens is fixed for given values of $\zL,\zS$ and $\mathbf{\Omega}$) these variables are depicted with a dot. The plates represent multiple independent systems with the same parametrization.}
\label{COSMIC_BEAMS_PGMs}
\end{figure}

Here we present a formulation to infer unbiased cosmological parameters from an impure sample of strong lenses. This was originally inspired by the zBEAMS \citep{Roberts2017_zBEAMS} and BEAMS \citep{Kunz2013,Hlozek2012,Kunz2007} methodologies but has been extensively adapted to the strong lensing case. We split the investigated science cases into 3 categories: Spectroscopic, Photometric and Photometric with contamination. The Probabilistic Graphical Models (PGM) for the latter scenario is shown in Figure \ref{COSMIC_BEAMS_PGMs}. We denote $\cosmo$ to be the cosmological parameters we wish to constrain $\cosmo = \{\Omega_m,\Omega_k,w_0,w_a\}$, and $\mathbf{D}$ to be the data. We wish to infer the posterior \posterior. 

For the cosmological inference, we adopt a Singular Isothermal Sphere (SIS) mass profile; isothermal mass profiles have been found to closely fit lens galaxies, a finding dubbed the `bulge-halo conspiracy', although small deviations have been found (e.g., \citealp{Auger2010_SLACS_X,Etherington2023_Beyond_the_bulgehalo_conspiracy}). 
In reality, the inferred cosmology is dependent on the value of the power-law slope, \gammalens{} (see e.g., \citealp{Li2024}). Such inference would benefit from space-based data (such as in the \emph{Euclid}-LSST overlapping regions) for which \gammalens{} would be easier to measure. Alternatively, possible \gammalens{} values could be marginalized over hierarchically (i.e., with a population mean and width), or at an individual level with the wide uncertainties driven by the ground-based data. We do not use such marginalization here and simply adopt an SIS model, however, we measure the precision with which \gammalens{} can be measured from LSST data in Sect. \ref{S:TP_modelling}.

In the isothermal case, the Einstein radius ($\theta_E$), velocity dispersion ($\sigma_v$), lens and source redshifts ($\zL$,$\zS$) are related by
\begin{equation}
    r\equiv \frac{\DLS(\zL,\zS,\cosmo)}{\DS(\zS,\cosmo)} = \frac{c^2\theta_E}{4\pi \sigma_v^2}
\end{equation}
The data vector for the $ith$ lens system, $D_i$, consists of the lens and source redshifts, (\zLobsi, \zSobsi), and the  ratio of Einstein radii/velocity dispersions (\robsi) defined above: $D_i = \{\zLobsi,\zSobsi,\robsi\}$.

\subsubsection{Spectroscopic Case}
In this work, we take `spectroscopic' to indicate perfect redshift measurements (i.e., with no uncertainty). While in practice spectroscopic measurements have a small uncertainty and may be more uncertain for the source than the lens, the redshift errors for the spectroscopic lens sample are insignificant ($1\%$) compared to the kinematic and $\theta_E$ measurements.

From Bayes' Theorem, the posterior from one system, $\posteriorone$ is proportional to\\
\begin{strip}
\rule{\dimexpr(0.5\textwidth-0.5\columnsep-0.4pt)}{0.4pt}%
\rule{0.4pt}{6pt}
\begin{equation}\label{Eq: Spectroscopic_Formulation}
\begin{split}
&\posteriorone \propto \prior\cdot P(D_i|\cosmo)
\\
&\propto \prior\cdot \int\int\int P(\robsi,\zLobsi,\zSobsi,\zLi,\zSi,\ri|\cosmo) \cdot dz_{L,i} dz_{S,i} dr_{i}
\\
&\propto \prior\cdot \int\int\int P(\robsi|\ri)P(\zLobsi|\zLi)P(\zSobsi|\zSi)P(\ri|\zLi,\zSi,\cosmo) P(\zLi)P(\zSi) \cdot dz_{L,i} dz_{S,i} dr_{i}\\
&\propto \prior\cdot P(\robsi|r(\zLtruei,\zStruei,\cosmo)) 
\end{split}
\end{equation}

where we have used the fact that in the spectroscopic case, 
$P(z_{{\rm{obs}},i}|z_i) = \delta(z_{{\rm{obs}},i}-z_i)$,
$P(\zi) = \delta(\zi-\ztruei)$  and that 
 $P(r|\zL,\zS,\mathbf{\Omega}) = \delta(r-r(\zL,\zS,\mathbf{\Omega}))$.
Given each lens system is independent, the combined posterior for $N$ systems in each of the scenarios considered here is given by $\posterior \propto \prior \prod_i^N{P(D_i|\cosmo)}$.  

\subsubsection{Photometric Case}
For the photometric scenario, we marginalize over the redshift distributions of each system. We treat the lens population hierarchically, modeling the lens redshift distribution as a 4-component Gaussian Mixture Model (GMM), with hyperparameters $\beta_L$, which could well fit the redshift distributions while minimizing the number of inferred hyperparameters. We modeled the lens-source redshift relation $P(\zSi|\zLi)$ with a redshift-dependent log-normal distribution, with hyperparameters $\mathbf{s} = \{\sigma_{c},\sigma_{m},q_{c},q_{m}\}$ which define a linear relation with respect to lens redshift ($q_i = q_c+\zLi \cdot q_m$, $\sigma_i = \sigma_c+\zLi \cdot \sigma_m$). The log-normal distribution is given by
\begin{equation}\label{Eqn: LogNorm zLzS Dependence}
    f(x,\mu,\sigma,q) = \frac{1}{q(x-\mu)\cdot\sqrt{2\pi}}\cdot\exp\left\{-\frac{1}{2q^2}\log^2\left(\frac{x-\mu}{\sigma}\right)\right\}
\end{equation}
where we define $x=\zS-\zL$ and fix $\mu=0$ which constrains $\zLi<\zSi$, as required.
The posterior for a single system is proportional to
\begin{equation}
    \begin{split}
    & \posteriorone \propto \prior\cdot P(D_i|\cosmo)\\ 
    & \propto \prior \int P(\robsi,\zLobsi,\zSobsi,\zLi,\zSi,\ri,\boldsymbol{\beta_L},\mathbf{s}|\cosmo)\cdot 
    d\zLi \, d\zSi \, d\ri \, d\boldsymbol{\beta_L} \, d\mathbf{s}\\ 
    & \propto \prior \int P(\robsi|\ri) P(\ri|\zLi,\zSi,\cosmo)P(\zLobsi|\zLi) P(\zSobsi|\zSi)P(\zLi|\boldsymbol{\beta_L}) P(\zSi|\zLi,\mathbf{s})P(\boldsymbol{\beta_L})P(\mathbf{s})\cdot d\zLi \, d\zSi \, d\ri \, d\boldsymbol{\beta_L} \, d\mathbf{s}\\
    & \propto \prior \int P(\robsi|r(\zLi,\zSi,\cosmo)) P(\zLobsi|\zLi) P(\zSobsi|\zSi)P(\zLi|\boldsymbol{\beta_L}) P(\zSi|\zLi,\mathbf{s})P(\boldsymbol{\beta_L})P(\mathbf{s}) \cdot d\zLi \,d\zSi \, d\boldsymbol{\beta_L} \, d\mathbf{s}\\
    \end{split}
\end{equation}
\subsubsection{Photometric Case, with Contamination}
Without spectroscopic confirmation of a lens system, it is possible (or even likely) that false positives may be included in a large sample of lens candidates, such as those anticipated from LSST and \emph{Euclid}. Without accounting for this possibility, the resulting posterior, \posterior, would be biased. To account for contamination, one requires probabilities that a given system is a lens, such as those discussed in \citet{Holloway2024}. We denote the lens population `\Lens', the non-lens (false positive) population `\Nonlens', and the binary variable denoting a lens (or not) $\tau$, i.e., $\tau=\Lens$ or $\tau=\Nonlens$. We define further parent hyperparameters for the false positive population, $\boldsymbol{\alpha},\boldsymbol{\beta},\boldsymbol{\gamma}$, to describe the distributions of $\robs$, $\zLobs$, and $\zSobs$ measurements respectively. These were each parameterized by 4-component GMMs giving sufficient flexibility to fit the range of distributions.
The posterior, \posteriorone, is proportional to:
\begin{equation}\label{Eq: Phot_with_contam}
    \begin{split}
    \posteriorone & \propto \prior\cdot P(D_i|\cosmo) \\
    & \propto \prior \sum_\tau P(\robsi,\zLobsi,\zSobsi, \tau|\cosmo)\\
    & \propto \prior \cdot \left[\left\{\PLi \cdot \int P(\robsi|r(\zLi,\zSi,\cosmo,\Lens))\right.\right.
    P(\zLobsi|\zLi) P(\zSobsi|\zSi)
    P(\zLi|\boldsymbol{\beta_L})\\    
    &\cdot P(\zSi|\zLi,\mathbf{s}) P(\mathbf{s})P(\boldsymbol{\beta_L}) 
     \cdot d\zLi \, d\zSi \, d\boldsymbol{\beta_L} \, d\mathbf{s} \bigg\} + \\
    &\left\{(1-\PLi) \cdot \int P(\robsi|\boldsymbol{\alpha}, {\Nonlens}) \right.
    \cdot P(\zLobsi | \boldsymbol{\beta},{\Nonlens}) 
    \cdot P(\zSobsi | \boldsymbol{\gamma}, {\Nonlens}) 
    \cdot P(\boldsymbol{\alpha}) P(\boldsymbol{\beta}) P(\boldsymbol{\gamma}) \cdot d\boldsymbol{\alpha} \, d\boldsymbol{\beta} \, d\boldsymbol{\gamma} \bigg\}\biggr]\\
    \end{split}
\end{equation}
\hfill\rule[-6pt]{0.4pt}{6.4pt}%
\rule{\dimexpr(0.5\textwidth-0.5\columnsep-1pt)}{0.4pt}
\end{strip}

where we have omitted steps identical to the previous scenarios and defined $P(\tau=\Lens)\equiv\PL$.

\subsubsection{Computational Efficiency}\label{S: Computational Efficiency}
Equation \ref{Eq: Phot_with_contam} includes marginalization over a large number of parameters, namely the true redshifts $\zL,\zS$ which scale with the number of systems, and hyperparameters modeling the lens and false positive populations ($\boldsymbol{\alpha}$, $\boldsymbol{\beta}$, $\boldsymbol{\gamma}$, $\boldsymbol{\beta_L}$, $\mathbf{s}$). For a realistic number of lens systems, this marginalization introduces significant memory costs. However, on the assumption that the measurements are independent, the posterior includes the likelihood product of the individual systems which could in theory be parallelized. For example, a dataset could be split into batches, for which samples of the per-batch posteriors are drawn. These could then be combined by using kernel density estimation (KDE) on these samples to obtain weighted samples of the final posterior. This process is non-trivial, as the KDE process is affected by `curse of dimensionality' with such a large number of hyperparameters. Therefore, we tested whether inferring the hyperparameters $\boldsymbol{\alpha}$,$\boldsymbol{\beta}$,$\boldsymbol{\gamma}$,$\boldsymbol{\beta_L}$,$\mathbf{s}$ separately before treating these as known constants in the cosmology inference would be a plausible solution. We found that such treatment still provides accurate posteriors.
Given that the shape of the posterior remained the same, and there was negligible change in the precision, we adopted this method for this work, splitting the photometric datasets into \rt{4} batches.

The integration over redshifts $\zL$,$\zS$ in Equation \ref{Eq: Phot_with_contam} cannot be done by MCMC sampling, since the posterior term involves the sum of two components $P(D_i|\mathbf{\Omega},\rm {Lens})$ and $P(D_i|\mathbf{\Omega},\rm {Non\mhyphen Lens})$, of which only one includes integration over the true redshift values. Unlike typical likelihood functions involving simply the product of terms which could simply be sampled from, the absolute values of the two components is important here as this governs the weighting of the lens and false positive terms. We therefore calculated this integral numerically via the trapezium rule.
\begin{table}
\centering
\begin{tabular}{ccc}
\toprule
Parameter&Distribution&Arguments\\
\hline
$\Omega_m$&$U$(min,max)&(0,1)\\
$\Omega_k$&$U$(min,max)&(0,1)\\
$w_0$&$U$(min,max)&(-3,1)\\
$\alpha$($\mu$)&$U$(min,max)&(0,10)\\
$\alpha$($\sigma$)&Log-$U$(min,max)&(0.001,2)\\
$\alpha$($w$)&Dirichlet($N_{comp}$)&4\\
$\beta$($\mu$)&$U$(min,max)&(0,2)\\
$\beta$($\sigma$)&Log-$U$(min,max)&(0.001,1)\\
$\beta$($w$)&Dirichlet($N_{comp}$)&4\\
$\gamma$($\mu$)&$U$(min,max)&(0,10)\\
$\gamma$($\sigma$)&Log-$U$(min,max)&(0.001,2)\\
$\gamma$($w$)&Dirichlet($N_{comp}$)&4\\
$\beta_L$($\mu$)&$U$(min,max)&(0,2)\\
$\beta_L$($\sigma$)&Log-$U$(min,max)&(0.001,1)\\
$\beta_L$($w$)&Dirichlet($N_{comp}$)&4\\
$q_c$&$U$(min,max)&(0.01,1)\\
$q_m$&$U$(min,max)&(-1,1)\\
$\sigma_c$&$U$(min,max)&(0.1,5) \\
$\sigma_m$&$U$(min,max)&(0,6)\\
\bottomrule
\end{tabular}
\caption{Priors used for inference. Uniform and Log-Uniform distributions are denoted $U$ and Log-$U$ respectively.}
\label{tab:Prior_Table}
\end{table}
To draw samples from the posterior,  we used the JAX\footnote{\url{https://github.com/jax-ml/jax}} based \texttt{numpyro} package \citep{Numpyro_1_2019,Numpyro_2_2019}, and the No-U-Turn Sampler (NUTS, \citealp{NUTS_Sampler_2014}). Table \ref{tab:Prior_Table} shows the priors used for the cosmological parameters and other hyperparameters.

\subsection{Generation of Inference Data Vectors}\label{S:Inference_Dataset_Generation}
For the purposes of inferring cosmological parameters we used a larger test set than used for testing the networks' performances, now including both lenses and false positives. The COSMIC-BEAMS formalism requires 5 measurements for each system: the Einstein radius, $\theta_E$, the lens and source redshifts $\zL$,$\zS$, the lens velocity dispersion $\sigma_v$, and the probability that the system is a strong lens, $P_{\rm{L}}$. In each case, we generated realistic measurements/uncertainties to reflect those obtainable in LSST data. Naturally, these differed between the systems designated as being spectroscopic compared to photometric.
We used a 50:50 lens/false positive dataset in the photometric sample. While this ratio would vary depending on the particular lens classifier(s) used and the classifier score threshold chosen, this is an illustrative and realistic level to demonstrate the methodology and the effect of impurities in the sample. We note in that in \citet{Holloway2024} a 50:50 purity threshold would give a low completeness ($\sim40\%$) however given subsequent developments in lens classification (e.g., \citealp{Canameras2024,Schuldt2025,Walmsley2025}), a more complete sample than this is likely to be achievable with LSST data.
\\\\
\textbf{Lens Systems:}\\
For LSST we expect to identify $\mathcal{O}(10^5)$ lenses. Here we emulated this larger sample by modeling the parameter space of the test set described in Section \ref{S:Data} via KDE, in particular over redshift, velocity dispersion, effective radius and Einstein radius parameters, and drew $100$k lens systems from this distribution. 

The redshift measurements differed between the photometric and spectroscopic systems. In the spectroscopic case, we treated the redshift uncertainties as negligible (Eq. \ref{Eq: Spectroscopic_Formulation}).
For the photometric systems, 
we used uncertainties of $0.02\cdot(1+z)$ based on the LSST Science Requirements Document \citep{LSST_SRD}, and investigated in \citet{Graham2018}.

We combined the observed velocity dispersion and Einstein radius to give a single cosmology-dependent term $r_{\rm{True}}$ as shown in Figure \ref{COSMIC_BEAMS_PGMs}. In the photometric case, we assigned an uncertainty to these measurements of $\sigma_r=0.4$ based on the typical uncertainty resulting from 1) the Einstein radii measurements of the \NNfour{} network, i.e., without lens subtraction, applied to the LSST test set and 2) the velocity dispersion which could be estimated using photometry alone. This was derived using the typical scatter in the Fundamental Plane, seen for typical lens galaxies from the SLACS survey \citep{Auger2010_SLACS_X}. For the spectroscopic systems we used $\sigma_r=0.085$ based on the uncertainty from the Einstein radii measurements of the \NNfour{} network and a $10\,\rm{km\,s^{-1}}$ uncertainty in the velocity dispersion, chosen to match the value in \citet{Li2024}.\\\\

\noindent\textbf{False Positive Systems:}\\
To generate a population of false positives (i.e. non-lenses), we used example false positives from multiple lens classifiers applied to Hyper-Suprime Cam (HSC, \citealp{Aihara2019,Aihara2022}) survey data,
in particular, classifiers used in \citealp{Holloway2024} \citep{Sonnenfeld2020_SW_HSC,Canameras2021,Jaelani2023,Ishida2025}. Being a ground-based wide-area survey, HSC was the closest available survey to LSST, and thus the false positives and lenses identified previously in this survey are likely to be similar to those in LSST.
These classifiers were trained/tested on HSC images (which, in the case of the Citizen Science (`CS') classifier, \citealp{Sonnenfeld2020_SW_HSC}) had the central galaxy light subtracted). In this work, we selected the top 0.1\% of objects from each classifier from the cross-matched catalog described in \citet{Holloway2024}, and from these identified those with an expert grade of 0, indicating the systems were not real lenses. $i$-band cutouts of these systems were downloaded from the HSC archive\footnote{\url{https://hsc-release.mtk.nao.ac.jp/doc/}}, and pixel-matched to the LSST pixel scale (0.2\arcsec). 
We used the performance of the \NNfour{} network on the false positives, in particular, measurements of their `Einstein radii', from these classifiers to generate a dataset of realistic measurements for false positives. For automated lens model methods such as NPE, the modeller will always produce lens measurement outputs such as $\theta_E$ regardless of whether the cutout it is shown contains a lens system or not. Any systematic differences in such measurements between the lens and false positive samples can then be used as part of the inference to distinguish lenses and false positives. Due to the similarity between HSC and LSST, such measurements should be representative of those obtained from automated lens modeling of LSST systems.

The inference data vectors also required redshift measurements. Strong lens searches are often `targeted', whereby color and/or magnitude cuts are applied to a galaxy catalog to identify early-type galaxies to which lens classifiers are then applied. To avoid confusion here we will refer to the target galaxy of such searches as the primary object, and any surrounding objects as a secondary object. These would refer to the lens and source galaxy respectively in the case of a true positive. We assumed that the primary objects for both true and false positives will have the same redshift and velocity dispersion distributions due to the targeted nature of typical lens searches. This assumption for redshifts would not hold for the secondary objects, where true positives would require $\zL<\zS$, but false positives would not. For the false positives, we assumed that a secondary object would still be present, be that spiral arms confused for lensed arcs, or neighboring objects which are not lensed. In practice, systems for which a secondary object could not be identified/analyzed would likely be removed from any subsequent analysis so this assumption is realistic. For this work, we assumed that for the false positives the primary and secondary object redshifts were uncorrelated. Therefore, for the secondary objects we drew redshift values randomly from the photo-z measurements in the HSC-DEEP galaxy catalog \citep{Aihara2022}, with the 5-yr LSST \emph{i}-band depth applied. \\ 
In all cases, we assumed the measurements were on average unbiased but imprecise, with representative uncertainties based on the literature. In particular, we assumed measurements of the velocity dispersion from the fundamental plane were correct to within realistic uncertainties described above. In all cases, `measurements' were drawn from Normal (or truncated-Normal where relevant) distributions, with mean and standard deviation given by the true value and relevant uncertainty respectively.

\subsubsection{Choice of Prior for $P_{\Lens}$}
The prior lens probability, $P_{\Lens}$ represents the initial confidence that each system is strongly lensed. This would be derived from strong lens finders and/or automated modeling, and is discussed further in Section \ref{S: Inference_with_Impure_Samples}. We find that the hyperparameters fixed prior to conducting the cosmological inference require a certain confidence to be determined accurately. The inference test-set consisted of $\sim50:50$ lenses:false positives, totaling \rt{$\sim200$\,k} systems. 
We assigned the photometric sample $P_{\Lens}$ values of $P_{\rm {thres}}$ and $1-P_{\rm {thres}}$ in equal proportions. We found $P_{\rm {thres}}\geq0.9$ was required for the fixed-hyperparameters discussed in Section \ref{S: Computational Efficiency} not to bias the resulting cosmological inference and thus fix \rt{$P_{\rm {thres}}=0.9$}. We did not update the $P_{\Lens}$ values during the inference (i.e., they were kept fixed), to ensure they remained accurately calibrated.
\section{Results}\label{S:Results}
\subsection{Lens Modeling of LSST Lenses}\label{S:TP_modelling}
\begin{figure*}
 \centering
  \centering
  \includegraphics[width=0.8\textwidth]{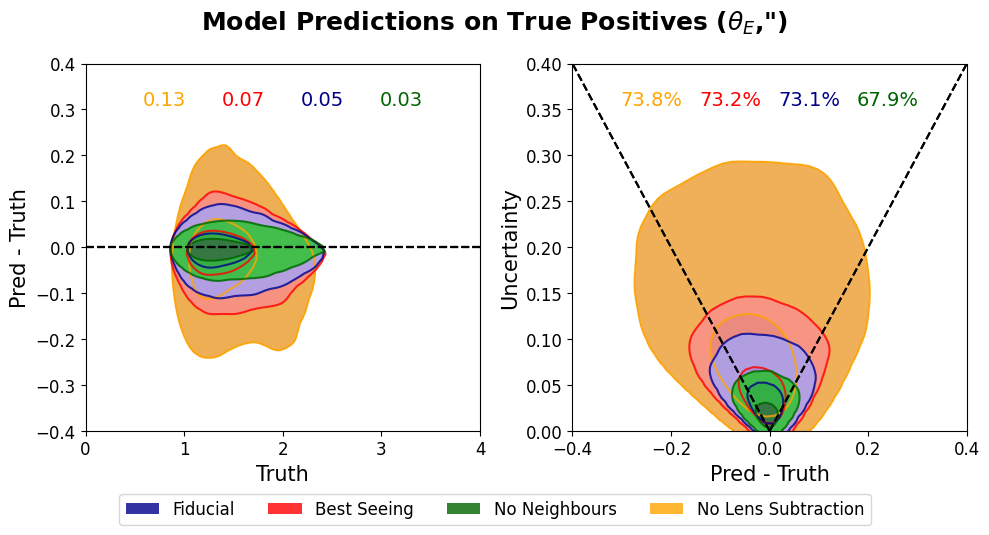}
\caption{A comparison of the performance of the neural networks when trained using different training images. The best performance is seen when not injecting into DP0.2 images, and the second best is seen from the Fiducial model using full-epoch coadds (rather than down-selecting the best-seeing single exposures). In general, the uncertainties produced by the networks are well calibrated. The precision values for each network are highlighted in the left-hand plot, while the proportion of systems with model predictions less than 1-sigma from the truth are listed in the right-hand plot.}
\label{Model_Performance_tE_Only}
\end{figure*}

We first present the results of our modeling networks on realistic simulated LSST lens systems. Figure \ref{Model_Performance_tE_Only} shows the network recovery of the Einstein radii in our test sets, along with the network uncertainties. We achieve a mean 
precision on $\theta_E$ of \rt{$3.7, 5.0, 2.2$ and $9.6\%$} for networks \NNone, \NNtwo, \NNthree{} and \NNfour{} respectively (defined in Sect. \ref{S:Network_Training}). The best performing network was the \NNthree{} network, which was the most simple, but least realistic, simulation. We find the systems with the highest precision are unsurprisingly those with the highest magnification and brightest source galaxies which produce more visible lensed arcs. We note that the \NNone{} network is $2.6\times$ more precise than the \NNfour{} network.  For the inference we use precision values based on the latter network as we use unsubtracted HSC cutouts for the false positive systems but highlight that the inference could be further improved by using systems with the central galaxy's light subtracted. We find each network's uncertainties are well calibrated, with \rt{$69.2-72.1\%$} of predicted $\theta_E$ values within $1\sigma$ of the ground truth. We provide the precision and mean bias achieved by each network for a range of parameters in Table \ref{T: Network_Precision}.  

\begin{table*}
\centering
\caption{Mean precision and bias values for each LSST realization. We find there is consistent benefit to performing lens subtraction prior to lens modeling, while the \NNtwo{} network typically performs slightly worse than the \NNone{} network. Note that all networks except that labeled as `No Lens Subtraction' use Poisson-limited lens subtracted images.}
\label{T: Network_Precision}
\begin{tabular}{llccc}
\toprule
Parameter & Network & Precision & Bias & Bias ($\sigma$) \\
\midrule
Einstein radius, $\theta_E$ (\arcsec) & Fiducial & 0.05 & -0.01 & -0.2 \\
 & Best Seeing & 0.07 & -0.02 & -0.3 \\
 & No Neighbours & 0.03 & -0.008 & -0.3 \\
 & No Lens Subtraction & 0.1 & -0.04 & -0.3 \\
  \cline{2-5}
Density Slope, $\gamma$ & Fiducial & 0.2 & 0.07 & 0.4 \\
 & Best Seeing & 0.2 & 0.07 & 0.3 \\
 & No Neighbours & 0.1 & 0.07 & 0.4 \\
 & No Lens Subtraction & 0.2 & 0.06 & 0.3 \\
   \cline{2-5}
Lens Shear,$\gamma_1$ & Fiducial & 0.04 & -0.001 & -0.01 \\
 & Best Seeing & 0.05 & -5e-4 & 2e-4 \\
 & No Neighbours & 0.03 & -8e-4 & -0.02 \\
 & No Lens Subtraction & 0.06 & -0.002 & -0.04 \\
   \cline{2-5}
Lens Ellipticity, $e_1$ & Fiducial & 0.08 & -0.001 & -0.004 \\
 & Best Seeing & 0.09 & -4e-4 & 0.003 \\
 & No Neighbours & 0.05 & -9e-4 & -0.009 \\
 & No Lens Subtraction & 0.1 & -0.007 & -0.06 \\
   \cline{2-5}
Lens Center, x (") & Fiducial & 0.06 & -0.001 & -0.02 \\
 & Best Seeing & 0.08 & 6e-4 & 0.02 \\
 & No Neighbours & 0.04 & 3e-4 & -0.003 \\
 & No Lens Subtraction & 0.003 & -1e-5 & 0.07 \\
\bottomrule
\end{tabular}
\end{table*}
While it may be expected that forming a coadd image of only the best seeing top $1/3$ of single exposures would improve the model precision, i.e. the \NNtwo{} network, we find that the precision is decreased compared to the \NNone{} case. This is likely because of the corresponding decrease in depth, of $\sim0.6$mag from reducing the number of single exposures contributing to the coadd image. However, we note that in practice this result may change depending the choice of `best-seeing' single exposures used to generate the coadd.

\begin{figure*}
\centering
\includegraphics[width=\textwidth]{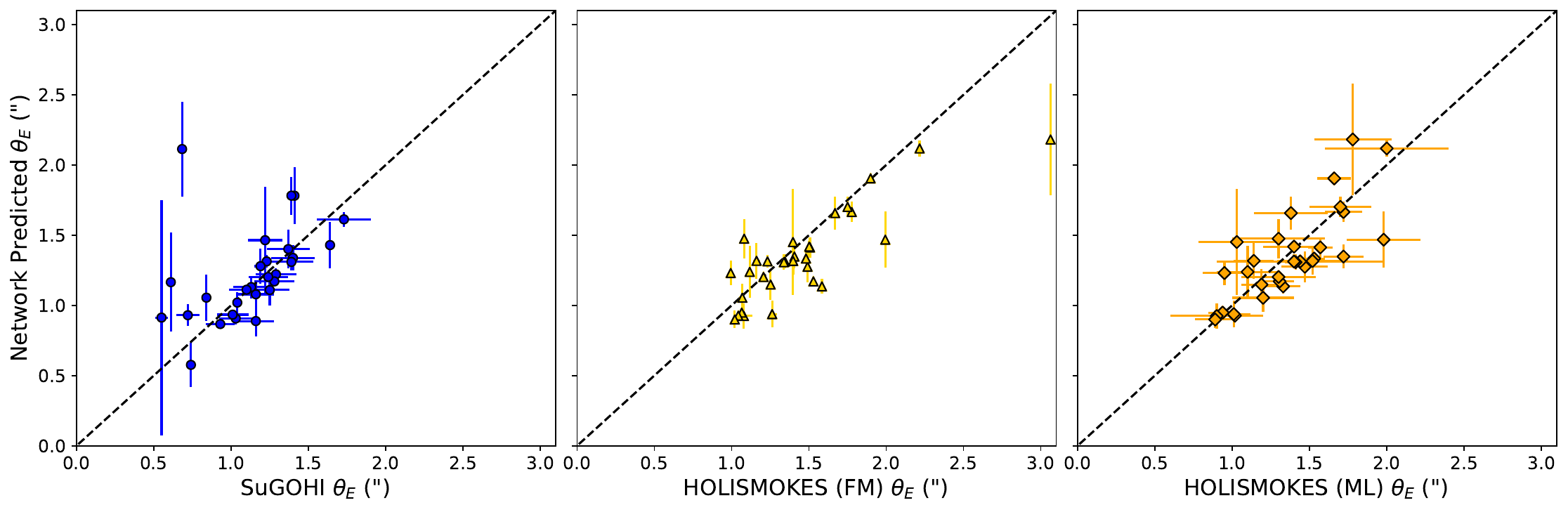}
\caption{Comparison of the $\theta_E$ values from the literature (SuGOHI catalogue and \citealp{Schuldt2023}) compared to the predictions of the \NNfour{} network. In the center and right right-hand panels we plot predictions from a machine learning ResNet model (`ML') and the forward-modelling code \texttt{GLEE \& GLAD} (`FM') from \citet{Schuldt2023}. In general we find good agreement between the \texttt{paltas} predictions and literature values, even though the network in this work was not trained on HSC data.}
\label{HSC_TP_Predictions}
\end{figure*}

We tested the \NNfour{} network on true A-grade lens systems from HSC which had measurements of $\theta_E$ available in the literature, having adjusted the cutout pixel scale of these HSC systems to that of LSST. We used $\theta_E$ measurements provided in the SuGOHI catalog\footnote{\url{http://www-utap.phys.s.u-tokyo.ac.jp/\~oguri/sugohi/} accessed 9/4/2024.}
and in HOLISMOKES X \citep{Schuldt2023} as comparison values. In general, there is good agreement as shown in Figure \ref{HSC_TP_Predictions} which compares the Einstein radii estimated by the \NNfour{} network with the values from the literature for these A-grade systems.

\subsection{Lens Modeling of False Positives}\label{S:FP_modelling}
We applied the \NNfour{} network to cutouts of false positives as described in Section \ref{S:Inference_Dataset_Generation} to determine the behavior of the modeler when shown objects that were not lenses.
\begin{figure*}
 \centering
  \centering
  \includegraphics[width=0.8\textwidth]{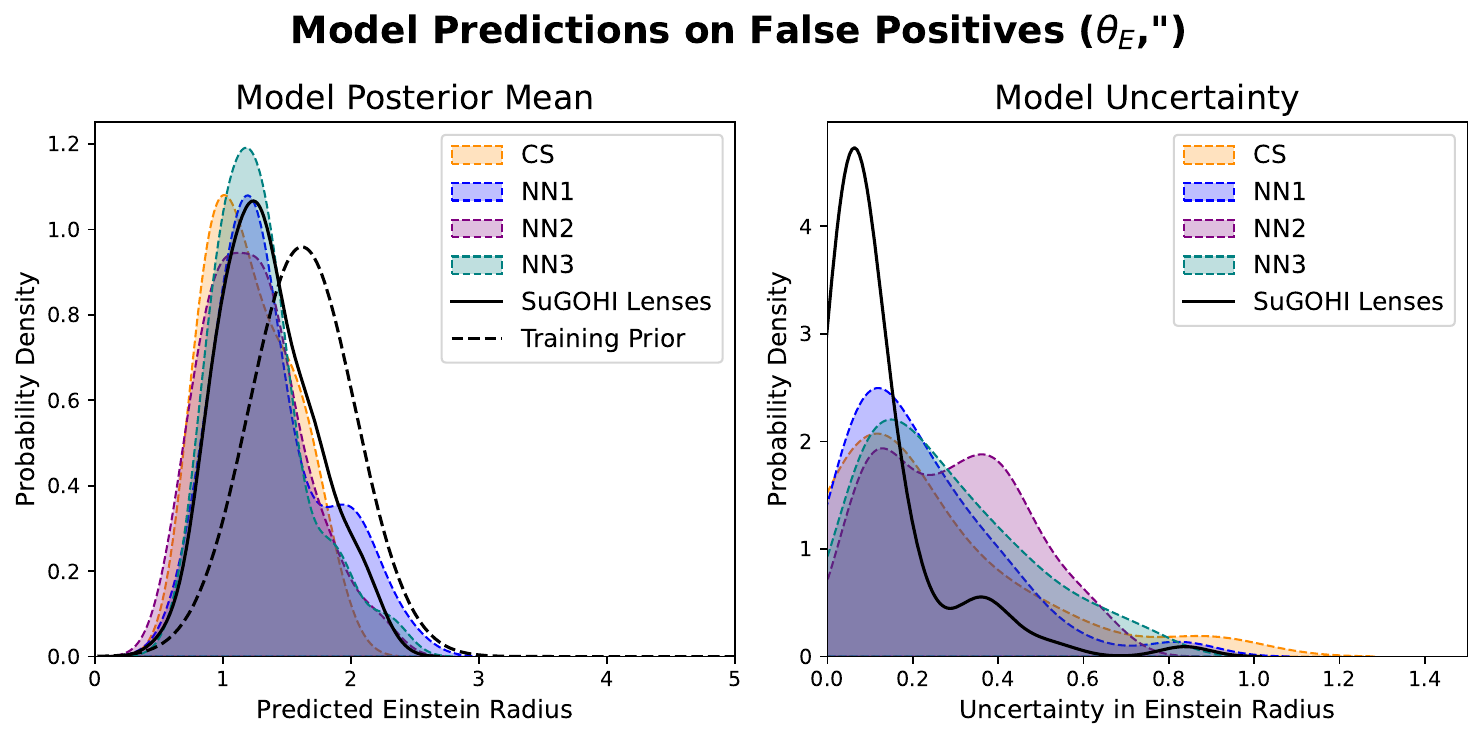}
\caption{Distribution of $\theta_E$ values predicted by the \NNfour{} network for a range of false positives from HSC searches. These false positives were drawn from the following searches: 
NN1: \citealp{Canameras2021}, 
NN2: \citealp{Ishida2025}, 
NN3: \citealp{Jaelani2023}, 
CS: \citealp{Sonnenfeld2020_SW_HSC}. The left-hand plot shows the distribution of the network output posterior means, and the right-hand plot shows the network uncertainty distribution. The solid-black curve shows the same distributions, but for A-grade lens candidates in HSC. The dashed curve (left) shows the distribution of Einstein radii from which the training set was drawn.}
\label{HSC_FP_Predictions}
\end{figure*}
Figure \ref{HSC_FP_Predictions} shows the distribution of network-predicted Einstein radii for the false positives compared to the distribution of predicted values for lens candidates from HSC. We find the $\theta_E$ distribution differs between the false positive and true lens systems, and also differs from the network training prior i.e., when shown a false positive the network does not default to the most likely solution from its training set. The network is significantly less confident about its estimate of $\theta_E$ when presented with a false positive compared with a lens system. Such differences continue when considering other parameters learned by the network, for example, the network typically predicted shallower mass-density profiles (\gammalens) for false positives, i.e. closer to a mass-sheet, than for true lenses as demonstrated in Appendix \ref{A: Network_Behaviour_to_All_Parameters}. 

\subsection{Posterior Images for Lenses and False Positives}\label{S: Posterior Images}
Given lens parameter posterior distributions, one can reconstruct `posterior images' by drawing lens parameters from these posteriors. To achieve this, we retrained the \NNfour{} network with additional parameters (size and magnitudes of the lens and source, and source position). We note this did not cause a significant change in uncertainty or error in the primary parameters of interest, $\theta_E$ and \gammalens. Figures \ref{Posterior_Predictions_TPs} and \ref{Posterior_Predictions_FPs} show such reconstructed images for a sample of the HSC lens candidates and false positives used in this work. We find that the reconstructions for false positive systems often follow the orientations of objects in the nearby environment, for example neighboring galaxies or spiral arms. Reconstructions of lens candidates from HSC replicate the true images well. In the case of a quad \rt{(Fig. \ref{Posterior_Predictions_TPs}, row 3, left column)}, individual draws of the lens parameters produced varied orientations of the lensed image, and the median reconstruction does not have four clear images. This is likely because the training set contained mostly doubly imaged systems and did not contain point sources. Cases of single or double arcs are generally well replicated in the reconstruction (e.g., top left system in Fig. \ref{Posterior_Predictions_TPs}), with scatter originating in the differences in magnification between posterior samples rather than in arc number or orientation.

\subsection{Cosmological Inference from an Impure sample of Strong Lenses}\label{S:CosmoInference}
\begin{figure*}
 \centering
  \centering
  \includegraphics[width=0.8\textwidth]{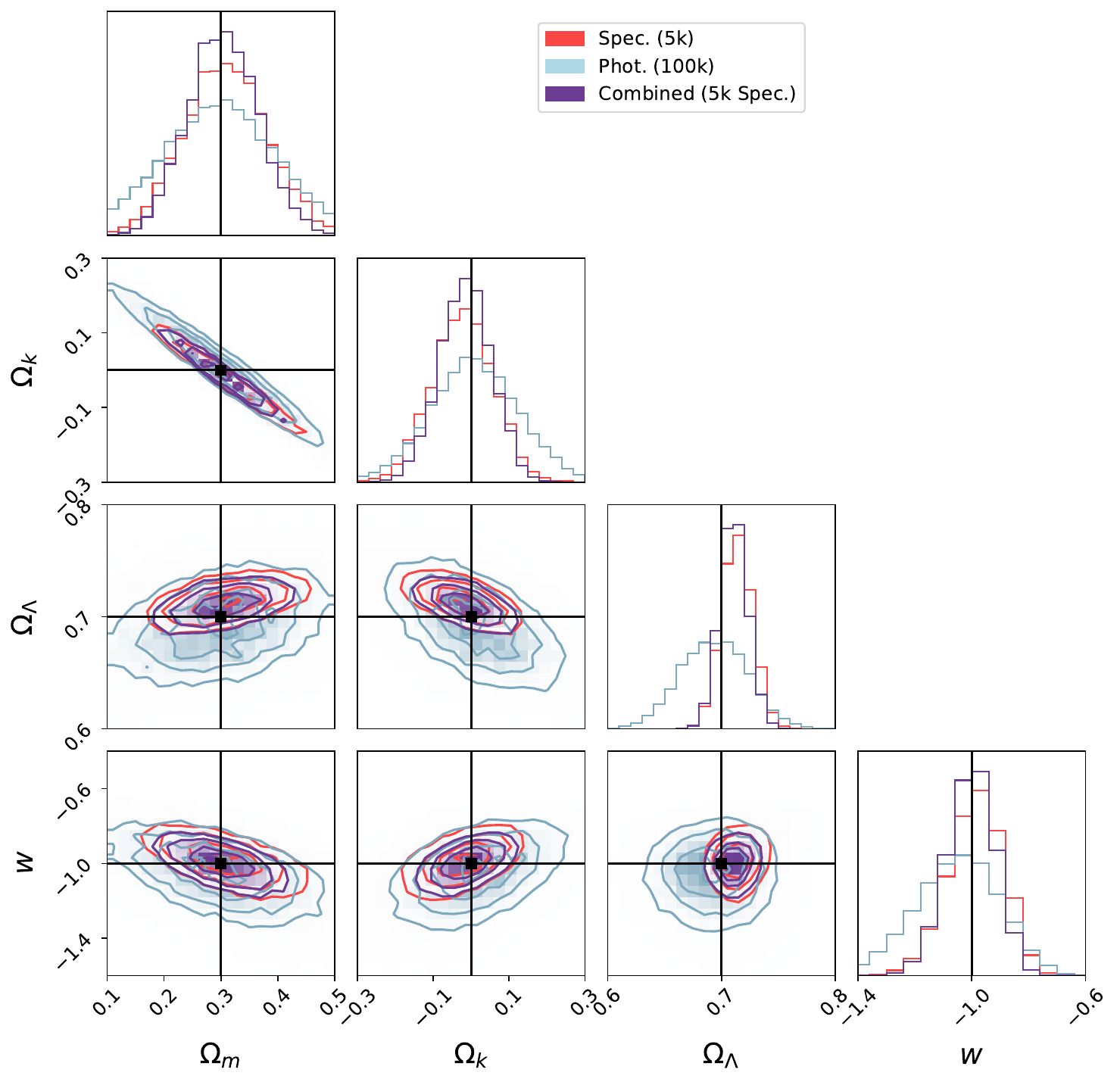}
\caption{Cosmological constraints for photometric (100k TP + 100k FP), spectroscopic (5k TP) and combined photometric + spectroscopic datasets. The ground truth cosmology ($\Omega_m=0.3,\Omega_k=0,w=-1$) is shown by the black lines. We find unbiased cosmological parameters can be inferred even with a contaminated dataset, and the cosmological precision improves when combining the photometric and spectroscopic datasets.}
\label{Cosmology_Posteriors}
\end{figure*}

The primary dataset for inferring cosmological parameters in this work is composed of both lens and false positive systems. 
Figure \ref{Cosmology_Posteriors} shows the cosmological posteriors obtainable from a sample of $100\,000+100\,000$ photometric lenses$+$false positives as well as $5000$ spectroscopic systems, and the constraints from combining these datasets. Posteriors from different samples from the inference test set are given in Appendix \ref{A: Cosmology Posterior Samples}. We find unbiased cosmological parameters can be obtained even with such significant contamination; the precision and biases found for a range of datasets are shown in Tables \ref{T: Cosmology_Precision} and \ref{T: Cosmology_Bias} respectively.

\begin{figure}
 \centering
\begin{subfigure}{0.5\textwidth}
  \includegraphics[width=\textwidth]{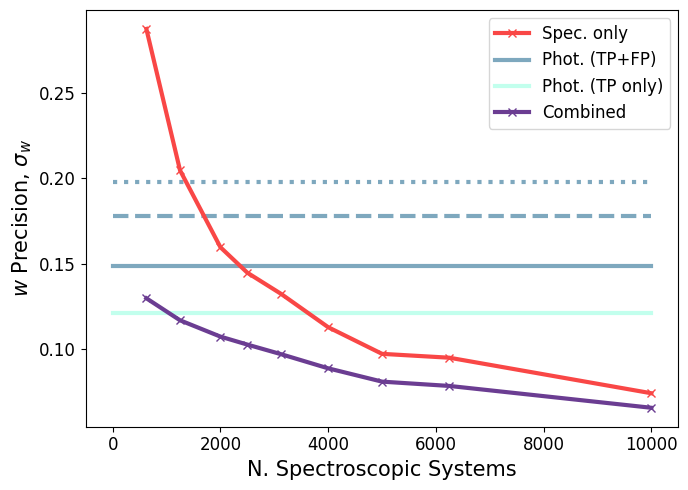}
\end{subfigure}
\caption{The median cosmological precision from a spectroscopic vs photometric dataset. The dotted, dashed and solid blue lines show the photometric sample precision from years 2, 5 and 10 respectively.}
\label{Spec_vs_Phot_precision}
\end{figure}

Figure \ref{Spec_vs_Phot_precision} shows a comparison of the precision in $w$ which can be obtained for a range of sizes ($625-10\,000$) of the spectroscopic dataset, which will depend on the fraction of systems for which source redshifts can be measured. We find that the contaminated photometric dataset provides roughly equivalent precision in $w$ to 2500 spectroscopic systems while the uncontaminated photometric dataset (i.e., with no false positives) provides improved precision, equivalent to 3500 spectroscopic systems. The precision increases with the larger photometric samples anticipated in later years of the survey. The combined photometric$+$spectroscopic precision improves upon the spectroscopic precision for all spectroscopic dataset sizes considered by $\Delta\sigma_w=0.05,\,0.02,\,0.01$ for $2$\,k, $5$\,k and $10$\,k spectroscopic systems respectively, with the greatest photometric benefit seen for $N_{\rm{spec}}\lesssim5000$. 

Cosmological parameter inference can be biased when the hierarchical model is not sufficiently flexible to accurately model the data. In this analysis we have assumed no evolution in the lens sample parameters such as in the mass density slope. This was analyzed in detail by \citet{Li2024} who found that assuming no evolution could bias the inferred cosmological parameters if evolution was present in the true sample. Therefore, such a possibility would need to be accounted for when applying this method to real data, for example, by inferring the scatter and evolution in the density slope simultaneously with the cosmological inference as in \citet{Li2024}.

\begin{table}
\centering
\caption{Median precision on cosmological parameters for different datasets and cosmological models.}
\label{T: Cosmology_Precision}
\addtolength{\tabcolsep}{-0.15em}
\begin{tabular}{llllll}
\toprule
$w$CDM & $\Omega_m$ & $\Omega_k$ & $\Omega_\Lambda$ &$w$&\\
\midrule
Phot. (TP-only) & 0.1 & 0.11 & 0.03 & 0.12 & / \\
Phot. (TP+FP) & 0.1 & 0.12 & 0.03 & 0.15 & / \\
Spec. (2k) & 0.11 & 0.13 & 0.02 & 0.16 & / \\
Spec. (5k) & 0.07 & 0.08 & 0.01 & 0.1 & / \\
Spec. (10k) & 0.05 & 0.06 & 0.01 & 0.07 & / \\
Spec. (2k) + Phot. (TP+FP) & 0.08 & 0.09 & 0.02 & 0.11 & / \\
Spec. (5k) + Phot. (TP+FP) & 0.06 & 0.07 & 0.01 & 0.08 & / \\
Spec. (10k) + Phot. (TP+FP) & 0.05 & 0.05 & 0.01 & 0.07 & / \\
 &  &  &  &  &  \\
 \toprule
$w_0w_a$CDM &  &  &  & $w_0$ & $w_a$ \\
\midrule
Phot. (TP-only) & 0.11 & 0.12 & 0.04 & 0.18 & 0.86 \\
Phot. (TP+FP) & 0.11 & 0.12 & 0.04 & 0.2 & 0.91 \\
Spec. (2k) & 0.12 & 0.13 & 0.05 & 0.22 & 1.0 \\
Spec. (5k) & 0.09 & 0.09 & 0.05 & 0.16 & 0.87 \\
Spec. (10k) & 0.08 & 0.06 & 0.05 & 0.12 & 0.76 \\
Spec. (2k) + Phot. (TP+FP) & 0.09 & 0.09 & 0.03 & 0.16 & 0.84 \\
Spec. (5k) + Phot. (TP+FP) & 0.08 & 0.07 & 0.03 & 0.14 & 0.81 \\
Spec. (10k) + Phot. (TP+FP) & 0.07 & 0.06 & 0.03 & 0.12 & 0.71 \\
\bottomrule
\end{tabular}
\end{table}

\begin{table}
\centering
\caption{Mean absolute bias, in units of $\sigma$ for the same range of datasets and cosmological models.}
\label{T: Cosmology_Bias}
\addtolength{\tabcolsep}{-0.15em}
\begin{tabular}{llllll}
\toprule
$w$CDM & $\Omega_m$ & $\Omega_k$ & $\Omega_\Lambda$ & $w$ & \\
\midrule
Phot. (TP-only) & 0.63 & 0.54 & 0.26 & 0.79 & / \\
Phot. (TP+FP) & 0.57 & 0.58 & 0.41 & 0.64 & / \\
Spec. (2k) & 0.81 & 0.78 & 0.64 & 0.93 & / \\
Spec. (5k) & 0.99 & 0.97 & 0.76 & 0.89 & / \\
Spec. (10k) & 0.71 & 0.66 & 0.52 & 0.60 & / \\
Spec. (2k) + Phot. (TP+FP) & 0.66 & 0.65 & 0.67 & 0.73 & / \\
Spec. (5k) + Phot. (TP+FP) & 0.83 & 0.83 & 0.77 & 0.77 & / \\
Spec. (10k) + Phot. (TP+FP) & 0.71 & 0.67 & 0.58 & 0.61 & / \\
 &  &  &  &  &  \\
\toprule
$w_0w_a$CDM &  &  &  & $w_0$ & $w_a$ \\
\midrule
Phot. (TP-only) & 0.98 & 0.96 & 0.16 & 0.79 & 0.58 \\
Phot. (TP+FP) & 0.60 & 0.59 & 0.55 & 0.75 & 0.71 \\
Spec. (2k) & 0.57 & 0.69 & 0.38 & 0.74 & 0.31 \\
Spec. (5k) & 0.76 & 0.93 & 0.48 & 0.87 & 0.54 \\
Spec. (10k) & 0.66 & 0.63 & 0.56 & 0.46 & 0.59 \\
Spec. (2k) + Phot. (TP+FP) & 0.57 & 0.67 & 0.49 & 0.73 & 0.81 \\
Spec. (5k) + Phot. (TP+FP) & 0.85 & 1.1 & 0.49 & 0.94 & 0.83 \\
Spec. (10k) + Phot. (TP+FP) & 0.60 & 0.61 & 0.56 & 0.69 & 0.65 \\
\bottomrule
\end{tabular}
\end{table}

\subsection{Effect of Bias in the Input Data}\label{S: Effect of Bias in the Input Data}
Throughout this work we have assumed that the input data is unbiased. Firstly the training set parameter distributions for the networks were centered on the test set distributions, reducing distribution shift. Secondly the input data vectors for the cosmological inference were uncertain, with non-negligible uncertainty, but unbiased on average. Finally, we assumed that the lens probabilities, \PL, were accurate. Here we discuss the consequences on the lens modeling and inference if these assumptions are not made.

The primary lens parameters of interest in this work were the \thetaE{} and \gammalens{}. The network bias for a parameter $p$ in the case where the training set distribution for the same parameter $p$ was shifted is shown in Table \ref{T: Network_Bias_from_Training}. In the case of \thetaE{} we find that the relative ($\sigma$) bias increases slightly but remains well below $1\sigma$ even for $10\%$ bias in \thetaE{}.
\begin{table}
\centering
\caption{The network bias in \thetaE{} and \gammalens{} which results from shifting the corresponding training set distributions away from the test-set mean values. }
\label{T: Network_Bias_from_Training}
\begin{tabular}{c c c c}
\toprule
Parameter, $p$&  Training&Model Bias&Model Bias ($\sigma$)\\
&  Bias in $p$ (\%)& &\\
\midrule
$\theta_E$ (") & -10 & -0.019 & -0.29 \\
& -1 & -0.015 & -0.26 \\
& 1 & -0.012 & -0.22 \\
& 10 & -0.012 & -0.30 \\
\gammalens{} & -10 & 0.007 & -0.02 \\
& -1 & 0.080 & 0.41 \\
& 1 & 0.080 & 0.39 \\
& 10 & 0.15 & 0.83 \\
\bottomrule
\end{tabular}
\end{table}
We find that the bias in \gammalens{} is more dependent on the training set distribution. In particular, increasing the \gammalens{} mean value in the training set produces a corresponding positive bias in the predicted values. We note that in NPE, such a training prior is usually divided out (see e.g, \citealp{WagnerCarena2023}) and replaced with a prior derived from the test set population which would likely reduce this bias. We do not perform this here since in reality, the photometric sample would include false positives and as described in Sect. \ref{S:FP_modelling}, the inferred parameter distributions for lenses/non-lenses differ. Therefore, the inferred distribution of \gammalens{} in a realistic photometric dataset would not necessarily match that of an uncontaminated lens sample and such reweighting could be biased. This could be overcome by using an uncontaminated sample of lenses (i.e with spectroscopic confirmation) on which to base this reweighting.

The cosmological bias due to biased input values of \PL{} and \robs{} is shown in Table \ref{T: Cosmology_Bias_with_Biased_Input}. These were generated by giving an additive bias of $\pm X$ for $X\in[0.01,0.05]$ in the case of \PL{}, or changing the input \robs{} values by a factor $1\pm Y$ for $Y\in[0.01,0.005,0.001]$. We find that the inferred cosmological values are correct to within $1\sigma$ for a \PL{} bias of $\pm 1\%$, but a $\lesssim0.5\%$ bias is required on \robs{}. In this work we have assumed such \robs{} measurements would be derived using the Fundamental Plane relation, and thus this would require careful calibration and comparison against spectroscopic measurements ($\sim10\%$ of the lens sample).

\citet{Kunz2007} discusses the possibility of including a bias term such that $\PLi{} \rightarrow \PLi{}+s_{\rm{bias}}$ in order to account for systematic shifts in \PLi{}, where this term is inferred alongside the main inference. In this case, inferring a large value of $s_{\rm{bias}}$ would serve as a warning of bias in the input catalog. We leave such analysis to future work but highlight that incorporating such a term could loosen the bias requirements discussed above.

\begin{table}
\centering
\caption{Mean absolute bias, in units of $\sigma$, after incorporating varying degrees of bias in the input parameters \PL{} and \robs{}, assuming a $w\rm{CDM}$ cosmology. The dataset used here was composed of 100k lenses and 100k non-lenses, corresponding to the Phot. (TP+FP) dataset in Table \ref{T: Cosmology_Bias} in the unbiased case.}
\label{T: Cosmology_Bias_with_Biased_Input}
\begin{tabular}{ccccc}
\toprule
\centering
 & $\Omega_m$ & $\Omega_k$ & $\Omega_\Lambda$ & $w$ \\
 \midrule
\PL{} (-5\%) & 2.3 & 2.4 & 1.3 & 2.8 \\
\PL{} (-1\%) & 0.56 & 0.51 & 0.28 & 0.89 \\
\PL{} (1\%) & 0.68 & 0.79 & 0.67 & 0.68 \\
\PL{} (5\%) & 3.2 & 3.2 & 1.3 & 2.5 \\
\robs{} (-1\%) & 0.28 & 0.76 & 1.7 & 1.2 \\
\robs{} (-0.5\%) & 0.30 & 0.49 & 1.1 & 0.76 \\
\robs{} (-0.1\%) & 0.25 & 0.32 & 0.53 & 0.61 \\
\robs{} (0.1\%) & 0.26 & 0.28 & 0.26 & 0.64 \\
\robs{} (0.5\%) & 0.23 & 0.28 & 0.47 & 0.87 \\
\robs{} (1\%) & 0.26 & 0.44 & 1.4 & 1.6 \\
\bottomrule
\end{tabular}
\end{table}
\section{Discussion}\label{S:Discussion}
\subsection{Modeling of LSST Lens Candidates}
In this work we trained and tested four different ML networks using realistic LSST-like simulations of $i$--band images. 
As shown by Table \ref{T: Network_Precision}, the trained networks can provide accurate measurements of a range of lens parameters, including the Einstein radius, the main focus of this work. The best performing network was the \NNthree{} network, which was also the most simple simulation. The small decrease in precision between \NNthree{} and \NNone{}, $\sigma_{\theta_E}=0.03\arcsec$ versus $\sigma_{\theta_E}=0.05\arcsec$, is indicative of the confusion of the network in identifying lensed sources in cutouts featuring neighboring objects. It is likely that such confusion would be reduced, and the overall precision improved,
by training the networks on multiple wavebands, such as shown by a $\sim20\%$ improvement in accuracy with multi-band modeling by \citet{Pearson2019}.

We assumed Poisson-limited lens subtraction, in the case of networks \NNone{}, \NNtwo{} and \NNthree{} to evaluate the benefits of removing the lens light for lens modeling. Lens subtraction can aid lens classification by reducing the effects of blending, and would likely reveal smaller $\theta_E$ systems in LSST images than would otherwise be detectable. While tools are available to do this automatically (e.g., \texttt{YATTALENS}, \citealp{Sonnenfeld2018}), imperfect lens subtraction could potentially bias the inferred lens model parameters. To account for such a possibility, networks trained for real LSST data may benefit from including residual lens light in the training images to ensure network resilience. Notably the \NNfour{} network could accurately measure the Einstein radius of grade A lens candidates from HSC (Fig. \ref{HSC_TP_Predictions}), even though it was not trained on HSC data, suggesting a degree of resilience for `easy' lens parameters. There was little agreement when compared to literature values for parameters such as ellipticity which may be more sensitive to variations in the PSF.

Beyond the Einstein radius, one parameter of interest was \gammalens, the mass density slope. This was fixed to $\gammalens=2$ (i.e., isothermal) in the test set following the \texttt{LensPop} population, but allowed to vary during network training. Given the seeing-limited imaging, \gammalens{} was a challenging parameter for the networks to learn, however the median precision in the \NNone{} network was substantially smaller than the training prior ($0.16$ versus $\sigma_{\rm{train}}=0.26$)
indicating it was possible to learn this parameter even in ground-based imaging. Furthermore, while the median precision for the \NNone{} network was $0.16$, this reduced to $0.09$ for the brightest $10\%$ of sources which would likely be those graded highest in a lens search.

One benefit of machine-learning lens modeling is speed and thus scalability. Following training ($\sim 10$ hours), producing models for $100\,000$ lens systems took $\sim17$ minutes on a CPU. Such rapid modeling, along with differences in model behavior between lenses and false positives (Figure \ref{HSC_FP_Predictions}), could allow for a `model-informed' classification of lens candidates to distinguish further true lenses from false positives. 

We investigated the failure modes of the \NNone{} network, in particular for which systems the network significantly over/underpredicted the Einstein radius. We found that over-predictions of the Einstein radius typically occurred when the unlensed source magnitude was very faint ($\sim27$) and the magnification was simultaneously low ($\sim3$). By contrast, under-predictions of the Einstein radius often occurred when the Einstein radius was very large, where the neighboring objects in the cutout from the DP0.2 simulation could easily be confused with the lensed source. Such outliers reduced by a factor of $3\times$ in the \NNthree{} network. However, overall the proportion of such outliers was very low, with \rt{$0.36\%$} of test subjects having a fractional error $>0.2$, and \rt{$0.18\%$} a fractional error $<-0.2$.

\subsection{Inference with Impure Samples of Strong Lenses}\label{S: Inference_with_Impure_Samples}
In this work we have demonstrated that unbiased cosmological parameters can be inferred even with a dataset contaminated by \rt{$50\%$} false positives and with photometric redshift uncertainties. 

Table \ref{T: Cosmology_Precision} shows the precision obtained for a range of datasets and cosmology models, with the mean absolute bias shown in Table \ref{T: Cosmology_Bias}. We note that the precision values obtained here are more optimistic than those identified by \citet{Li2024} for a spectroscopic dataset of the same size. This likely derives from the difference in mass model (isothermal vs a power-law profile including anisotropy) utilized by each method, as well as differences in $\theta_E$ precision expected between LSST (this work) and  \emph{Euclid} \citep{Li2024}. Nevertheless, the achievable precision from the photometric+spectroscopic dataset ($\sigma_w=0.07$ for $N_{\rm{spec}}=10\rm{\,k}$) is broadly comparable to that from the recent (Flat)-$w$CDM results of 
DESI ($\sigma_w=0.078$,  \citealp{DESI_II_2025}), 
eBOSS ($\sigma_w=0.15$, \citealp{Alam2021}) and
DES Year 5 supernovae results ($\sigma_w=0.15$, \citealp{DES2024}).
Furthermore, the combined LSST probes are expected to provide significantly tighter constraints than the individual probes \citep{DESC2018,Shajib2024}. A complete forecast of the combined strong lensing probes will be undertaken in future work (in prep.). 

The photometric sample provides the greatest improvements in $w$-precision for $N_{\rm{spec}}\lesssim5000$. Without spectroscopy, the Year-2 lens sample would provide a precision of $\sigma_w=0.2$ (again highlighting the above caveats regarding mass model assumptions), and the full 10-year photometric lens sample would be required for such constraints to be competitive with other probes (Figure \ref{Spec_vs_Phot_precision}). The difference in precision between the contaminated and uncontaminated photometric datasets ($\sigma_w=0.15$ versus $\sigma_w=0.12$) demonstrates the case for continuing to improve lens classification methods, as increased purity (for the same number of lenses) materially improves the precision. The number of systems for which spectroscopic data is available will depend on a number of factors, including the proportion of lenses detectable in LSST versus those identified in \emph{Euclid}. Lenses located in the overlapping footprint between \emph{Euclid} and LSST will be of high interest. The greater image quality of \emph{Euclid} VIS band will improve the lens modeling precision, though this was not the greatest source of uncertainty in this work, while also helping remove false positives from the photometric lens candidate sample. The \emph{Euclid} data could also help identify smaller $\theta_E$ systems which could otherwise be ambiguous candidates with LSST-alone. Finally, these overlapping lens candidates will also benefit from the improved photometry (and photo-z's) from the LSST optical bands and \emph{Euclid} NIR bands.

While cosmology has been the primary focus, the inference method presented here could be adapted easily to other analyses of the strong lens population. This work involved two primary sources of uncertainty: the velocity dispersion and photometric redshifts. Investigations that did not require either or both of these would gain more significantly from the inclusion of the photometric sample. For example, inferring the evolution of the power-law slope, which may be best suited to the overlapping data from \emph{Euclid} and LSST, would not require an estimate of the velocity dispersion. This has been discussed at length in the literature (e.g., \citealp{Ruff2011,Sonnenfeld2013,Xu2017,Remus2017}), and would benefit from the anticipated large photometric dataset from these surveys. We leave such investigations to future work.

Given the size of the photometric dataset, we find any inference is sensitive to small biases in the assumed parent hyperparameters, or correlations in the dataset. The inference presented here makes some simplifying assumptions, for example combining $\theta_E$ and $\sigma_v$ into a single $r_{true}$  with fixed uncertainty and fixing the parent hyperparameters prior to cosmological inference, which would need to be managed carefully when applied to real data. However, this work demonstrates that contamination by false positives is a tractable problem with significant potential for population inference. 

One assumption of the COSMIC-BEAMS formulation has been the availability of probabilities that each system is a strong lens. Extremes of $P_{\Lens}$ values would be easily obtainable either from the 4SLSLS survey for confirmed lenses, or expert-inspected false positives. Interim values could be derived from expert-graded (or citizen-graded, see \citealp{Holloway2025_EuclidQ1}) datasets, using the fraction of these confirmed (or rejected) by spectroscopic follow-up to map these grades to true probabilities that each system is a lens. A broader dataset of expert-graded systems could be obtained using available strong lens databases (e.g., SLED\footnote{\url{https://sled.amnh.org/}} and masterlens\footnote{\url{ http://admin.masterlens.org/index.php}}, \citealp{Moustakas2012}), which could be used to calibrate lens classifiers to map their outputs to lens probabilities at scale \citep{Holloway2024}.

\section{Conclusion}\label{S:Conclusion}
In this work we have detailed a formulation to infer cosmological parameters from an impure sample of strong gravitational lenses. In addition, we have demonstrated the precision with which lens parameters can be measured at LSST-scale using Neural Posterior Estimation. Our conclusions are as follows:
\begin{itemize}
    \item \textbf{Lens Parameter Precision:} We find that for our Fiducial (lens-subtracted) dataset Einstein radii can be measured to a precision of \rt{$3.6\%$}. The most accurate measurements were obtained from systems with high magnification and bright source galaxies. Furthermore, model precision was optimised when using full coadd images, rather than when using coadds generated using only single-exposures with the best seeing.
    \item \textbf{Behaviour of NPE network to False Positives:} We find our lens-modeling neural network is more uncertain when presented with systems that are not lenses. Given the rapid speed with which NPE can estimate lens parameters, such differences in behavior between true lenses and false positives could become a useful tool for automated lens classification, sifting likely lenses from false positives based on their measured lens parameters. For the `Einstein radius' measurement, the network commonly mis-identifies alternative light sources in the cutout to be the Einstein ring when applied to systems which are not strong lenses.
    \item \textbf{Cosmological precision from an impure sample of strong lenses:} We find unbiased cosmology can be inferred from an impure sample of strong lenses when the false positive population is accurately accounted for (i.e., via calibrated lens probabilities). We find the following $w$-precision would be obtainable for LSST lens candidate systems under $w$CDM and the assumption of an isothermal mass profile:
    \begin{itemize}
        \item 100k phot. (TP-only): $\sigma_w=\rt{0.12}$
        \item 100k+100k phot. (TP+FP): $\sigma_w=\rt{0.15}$
        \item 100k+100k phot. + 10k spec.: $\sigma_w=\rt{0.07}$
    \end{itemize}
    The photometric sample of lenses is comparable to 2500-3500 spectroscopic systems depending on sample purity, and provides significant improvement in $w$-precision over the spectroscopic sample alone up to $N_{\rm{spec}}\sim5\,000$ systems.
    \item The primary source of uncertainty in this work derives from the velocity dispersion which for the photometric dataset is assumed to come from the scatter in the Fundamental Plane. Given the large photometric dataset anticipated from LSST (and similarly \emph{Euclid}), population analysis that does not require a measure of the velocity dispersion would benefit significantly from this dataset, even with contamination up to $50\%$.
\end{itemize}
\section{Data Availability}
The code repository for \texttt{COSMIC-BEAMS} is located at \url{https://github.com/P-1884/COSMIC-BEAMS}. Additional data may be made available upon request to the corresponding author.

Full details of the sources of the false positive systems used in this work are described in \citet{Holloway2024}. The test set population described in Section \ref{S:Data} was generated using \texttt{LensPop}\footnote{\url{https://github.com/tcollett/LensPop}} \citep{Collett2015}, injected into DP0.2 data \citep{Korytov2019,DESC2021} using the \texttt{SLSim} LSST Science Pipeline\footnote{\url{https://github.com/LSST-strong-lensing/slsim}} \citep{Khadka2026}, using \texttt{Lenstronomy}\footnote{\url{https://github.com/lenstronomy/lenstronomy}} \citep{BirrerAmara2018,Birrer2021} for the lens simulation, and subsequently modeled via \texttt{paltas}\footnote{\url{https://github.com/swagnercarena/paltas}} \citep{WagnerCarena2023}. 
\section{Acknowledgements}
This paper has undergone internal review in the LSST
Dark Energy Science Collaboration. The internal reviewers were Simon Birrer and Renée Hložek.

PH led all aspects of the analysis. AV provided extensive feedback as well as manuscript editing throughout. PM, PV, SE and TL provided feedback on the cosmological inference and network training. SD provided feedback on the manuscript and TC aided the development of the LensPop test set. SB 
provided extensive feedback on the manuscript and analysis. 

PH thanks the DESC internal reviewers, Renée Hložek and Simon Birrer for their very helpful comments which improved the paper. 
PH acknowledges travel support provided by STFC for UK participation in LSST through grant ST/S006206/1. PH also acknowledges funding from the Science and Technology Facilities Council, grant code ST/W507726/1. AV acknowledges support from the Science and Technology Facilities Council, grant code ST/S006168/1 and ST/X00127X/1.

The DESC acknowledges ongoing support from the Institut National de 
Physique Nucl\'eaire et de Physique des Particules in France; the 
Science \& Technology Facilities Council in the United Kingdom; and the
Department of Energy and the LSST Discovery Alliance
in the United States.  DESC uses resources of the IN2P3 
Computing Center (CC-IN2P3--Lyon/Villeurbanne - France) funded by the 
Centre National de la Recherche Scientifique; the National Energy 
Research Scientific Computing Center, a DOE Office of Science User 
Facility supported by the Office of Science of the U.S.\ Department of
Energy under Contract No.\ DE-AC02-05CH11231; STFC DiRAC HPC Facilities, 
funded by UK BEIS National E-infrastructure capital grants; and the UK 
particle physics grid, supported by the GridPP Collaboration.  This 
work was performed in part under DOE Contract DE-AC02-76SF00515.

This work made use of data and computing resources provided by Vera C. Rubin Observatory, which is supported in part by the National Science Foundation through Cooperative Agreement AST-1258333 and Cooperative Support Agreement AST-1202910 managed by the Association of Universities for Research in Astronomy (AURA), and by the Department of Energy under Contract No. DE-AC02-76SF00515 with the SLAC National Accelerator Laboratory managed by Stanford University. Additional Rubin Observatory funding comes from private donations, grants to universities, and in-kind support from LSST-DA Institutional Members.

The Hyper Suprime-Cam (HSC) collaboration includes the astronomical communities of Japan and Taiwan, and Princeton University. The HSC instrumentation and software were developed by the National Astronomical Observatory of Japan (NAOJ), the Kavli Institute for the Physics and Mathematics of the Universe (Kavli IPMU), the University of Tokyo, the High Energy Accelerator Research Organization (KEK), the Academia Sinica Institute for Astronomy and Astrophysics in Taiwan (ASIAA), and Princeton University. Funding was contributed by the FIRST program from the Japanese Cabinet Office, the Ministry of Education, Culture, Sports, Science and Technology (MEXT), the Japan Society for the Promotion of Science (JSPS), Japan Science and Technology Agency (JST), the Toray Science Foundation, NAOJ, Kavli IPMU, KEK, ASIAA, and Princeton University.  This paper makes use of software developed for Vera C. Rubin Observatory. We thank the Rubin Observatory for making their code available as free software at http://pipelines.lsst.io/. This paper is based on data collected at the Subaru Telescope and retrieved from the HSC data archive system, which is operated by the Subaru Telescope and Astronomy Data Center (ADC) at NAOJ. Data analysis was in part carried out with the cooperation of Center for Computational Astrophysics (CfCA), NAOJ. We are honored and grateful for the opportunity of observing the Universe from Maunakea, which has the cultural, historical and natural significance in Hawaii. 
\bibliographystyle{mnras}
\bibliography{bibliography}
\appendix
\section{Posteriors for all Lens Parameters}
\begin{figure*}
 \centering
  \centering
  \includegraphics[height=0.95\textheight]{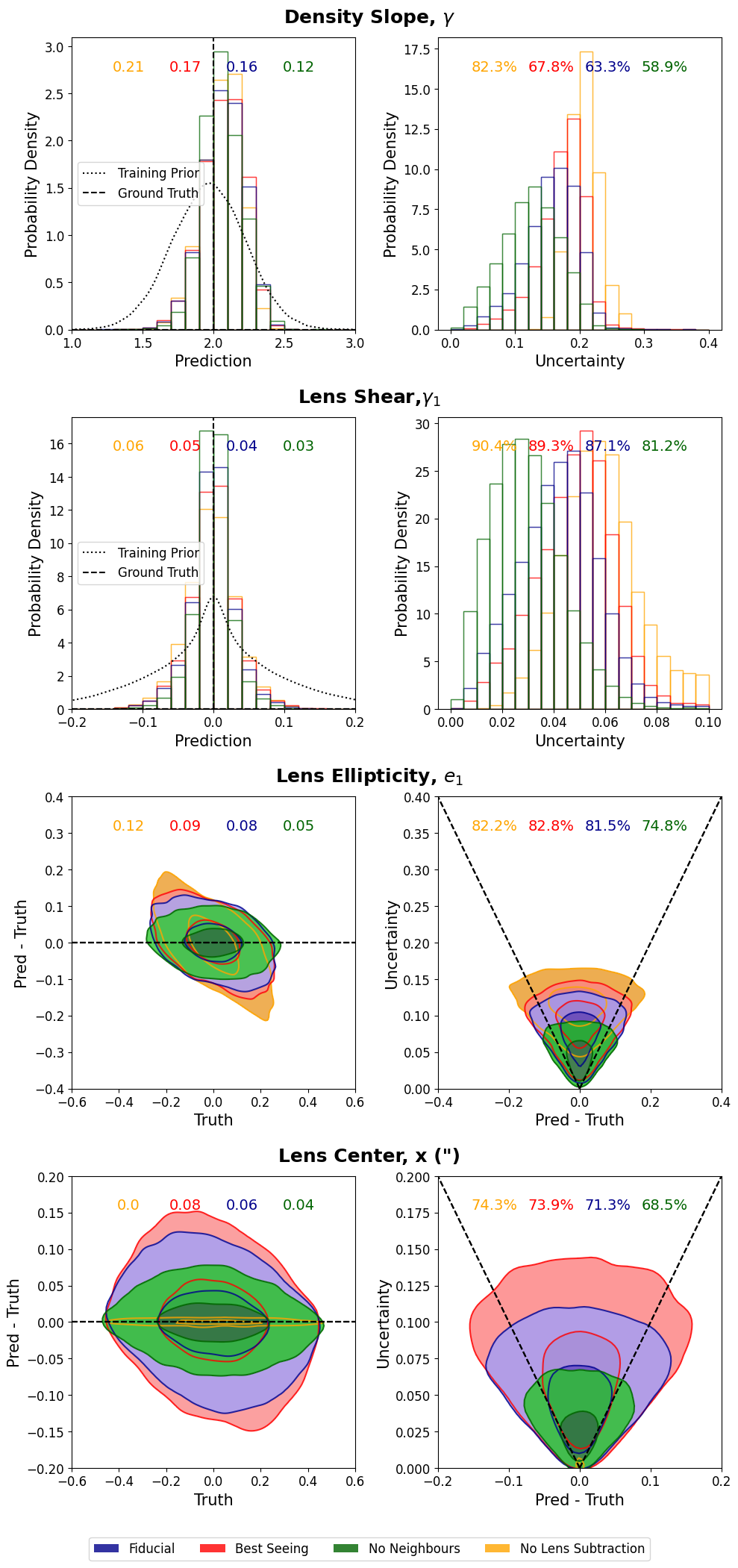}
\caption{Model performance for each of the networks across the other learned parameters. The best performing network is typically usually \NNthree{} however this uses the simplest test simulations. The improvement of \NNone{} over \NNfour{} demonstrates the benefit of lens subtracted images, which will also aid in lens finding. The precision values for each network are highlighted in the left-hand plots, while the proportion of systems with model predictions less than 1-sigma from the truth are listed in the right-hand plots.}
\label{Model_Performance_All}
\end{figure*}
Figure \ref{Model_Performance_All} shows the performance of each network to the remaining lens parameters. Histograms are plotted in the case of \gammalens{} and shear, for which the test set systems adopted a single value ($\gammalens=2,\gamma=0$). The \NNthree{} network consistently has the highest performance except in the case of learning the lens position, in which the \NNfour{} network unsurprisingly performs better. This demonstrates the importance of using realistic training/test sets, as oversimplified test sets (as known to be the case for \NNthree{}) can produce over-optimistic results. The \NNone{} network consistently performs better than the \NNtwo{} and \NNfour{} networks, demonstrating that signal-to-noise (from greater depth) and lens subtraction are important considerations for optimizing modeling precision.
\section{Network Behavior to False Positives across Lens Parameters}\label{A: Network_Behaviour_to_All_Parameters}
\begin{figure*}
\centering
\begin{subfigure}{0.8\textwidth}
  \centering
  \includegraphics[width=\textwidth]{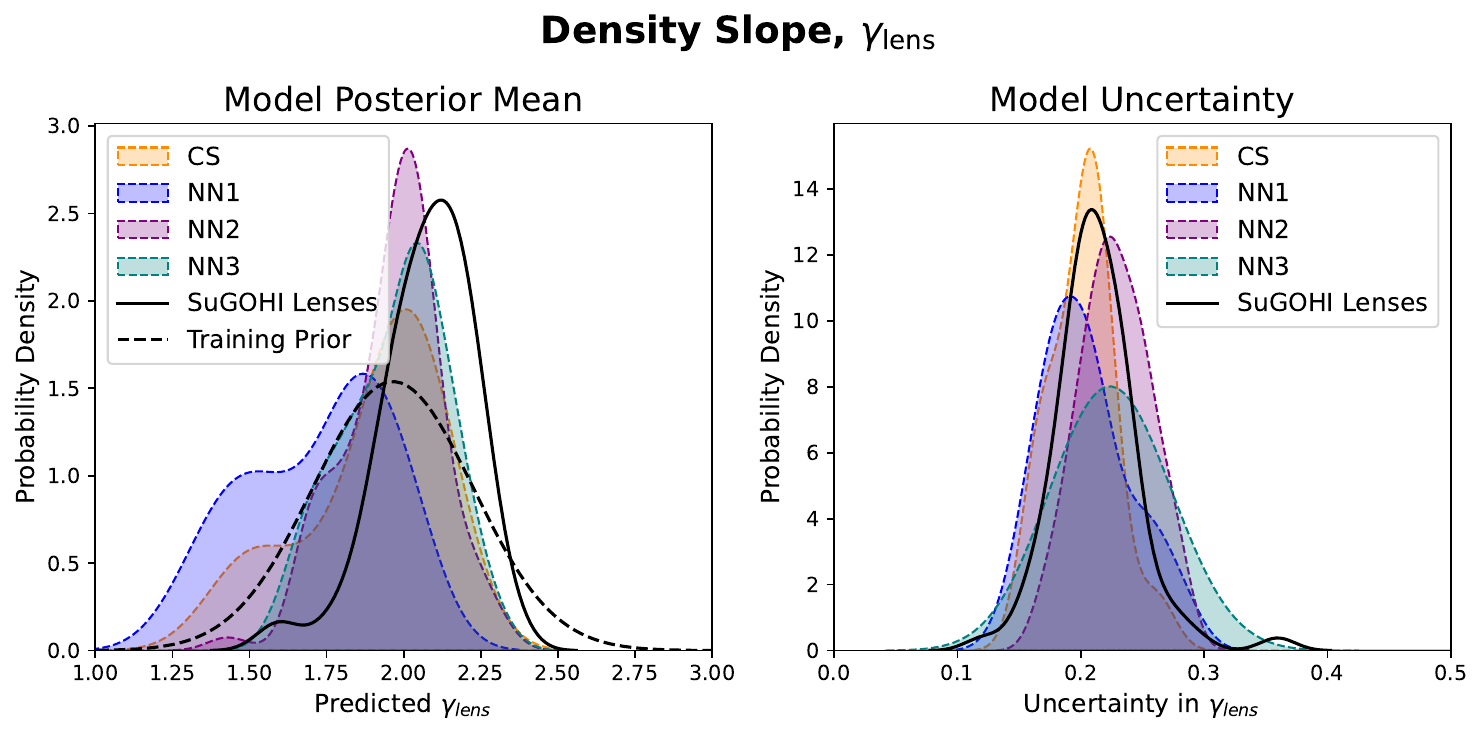}
\end{subfigure}
\begin{subfigure}{0.8\textwidth}
  \centering
  \includegraphics[width=\textwidth]{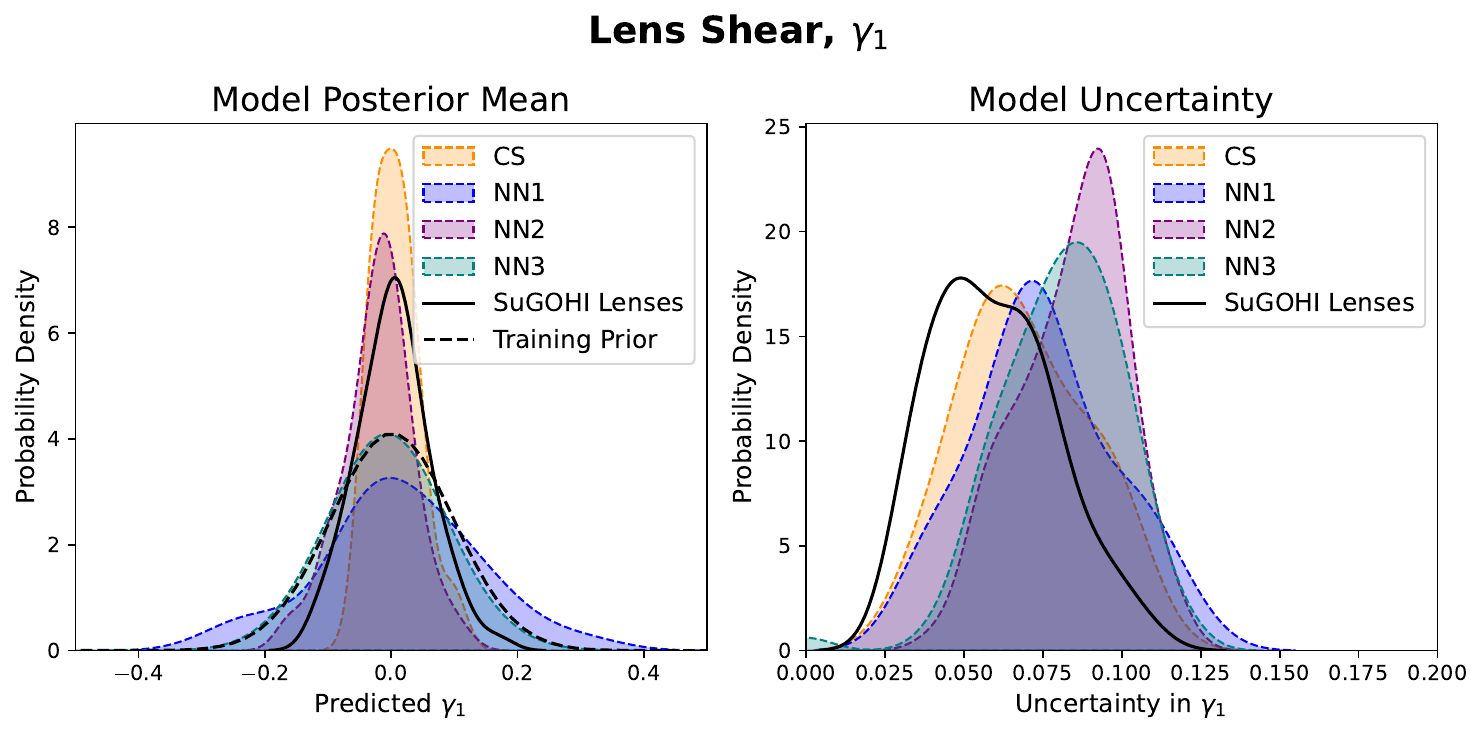}
\end{subfigure}
\begin{subfigure}{0.8\textwidth}
  \centering
  \includegraphics[width=\textwidth]{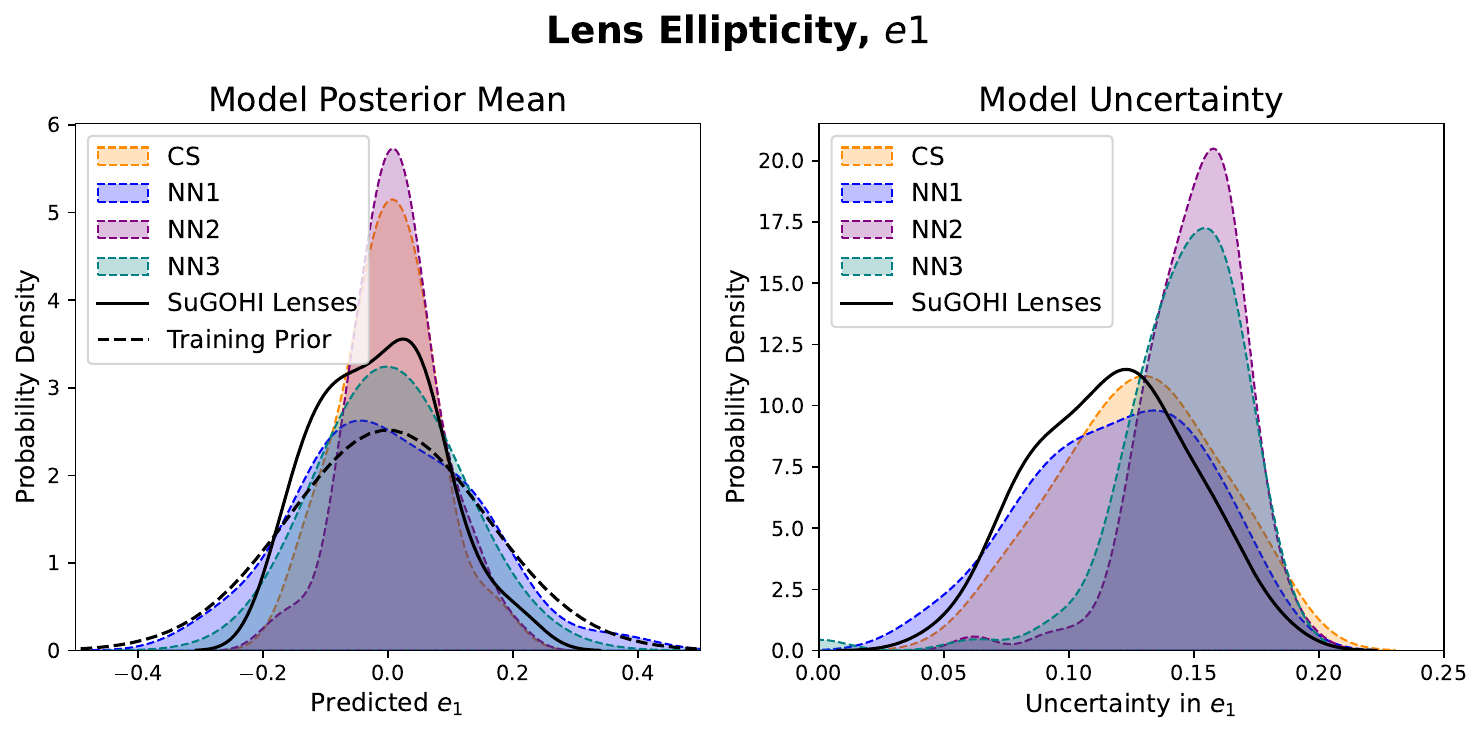}
\end{subfigure}
\caption{Comparison of the response of the \NNfour{} Network to false positives and true lens candidates for additional lens parameters. The false positives were taken from the following searches: 
NN1: \citealp{Canameras2021}, 
NN2: \citealp{Ishida2025}, 
NN3: \citealp{Jaelani2023}, 
CS: \citealp{Sonnenfeld2020_SW_HSC}. The distributions of the parameters of the A-grade SuGOHI lenses and training set systems are shown by the solid and dashed black lines respectively.}
\label{FP_Behaviour_All}
\end{figure*}

Figure \ref{FP_Behaviour_All} shows the distribution of lens model parameters predicted by the \NNfour{} network for false positives in comparison to true lenses. Most notable is the difference in predicted \gammalens{} values; the network predicts lower \gammalens{} values for the false positives than the SuGOHI lens sample. This could be because the network cannot identify lensing features in the false positive cutouts (as expected), and thus reduces the power-law slope to be closer to that of a non-lens (a $\gammalens$ value of $1$ would indicate a mass-sheet, which would not lens a background source). There is also greater uncertainty in the lens shear for the false positives but the remaining distributions are similar to those of the true lenses. 

\section{Cosmological Precision with different datasets}\label{A: Cosmology Posterior Samples}
\begin{figure*}
 \centering
  \centering
  \includegraphics[width=0.8\textwidth]{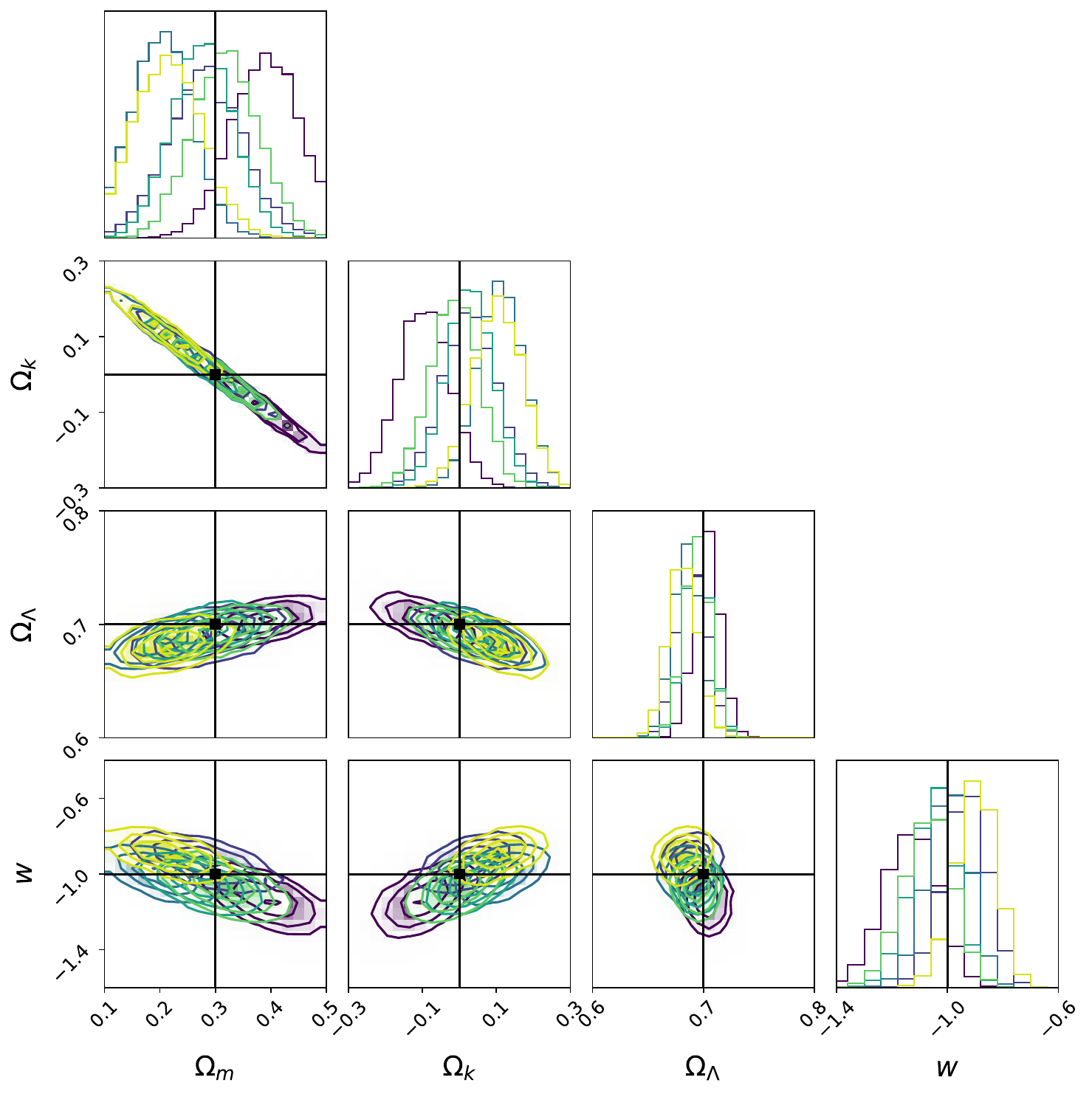}
\caption{Cosmological constraints from different random datasets of combined photometric (100k TP + 100k FP) + spectroscopic (5k) lenses. The ground-truth cosmology ($\Omega_m=0.3,\Omega_k=0,w=-1$) is shown by the black lines.}
\label{Cosmology_Posteriors_Samples}
\end{figure*}

Figure \ref{Cosmology_Posteriors_Samples} shows the cosmology posteriors for multiple photometric+spectroscopic datasets. These datasets were generated by drawing different samples of photometric ($100\,\rm{k}+100\,\rm{k}$ TP+FP) and spectroscopic ($5\,\rm{k}$) systems from the inference test set distribution. While there is relatively large scatter between the posteriors, on average they are unbiased. The mean bias values are given in Table \ref{T: Cosmology_Bias}. It is possible that a more complex mass model (or inclusion of evolution in the power-law slope) would increase the uncertainties in these posteriors, as discussed in Sections \ref{S:CosmoInference} and \ref{S: Inference_with_Impure_Samples}.

\section{Predicted Posterior Images from \texttt{paltas}}
\begin{figure*}
 \centering
  \centering
  \includegraphics[width=0.95\textwidth]{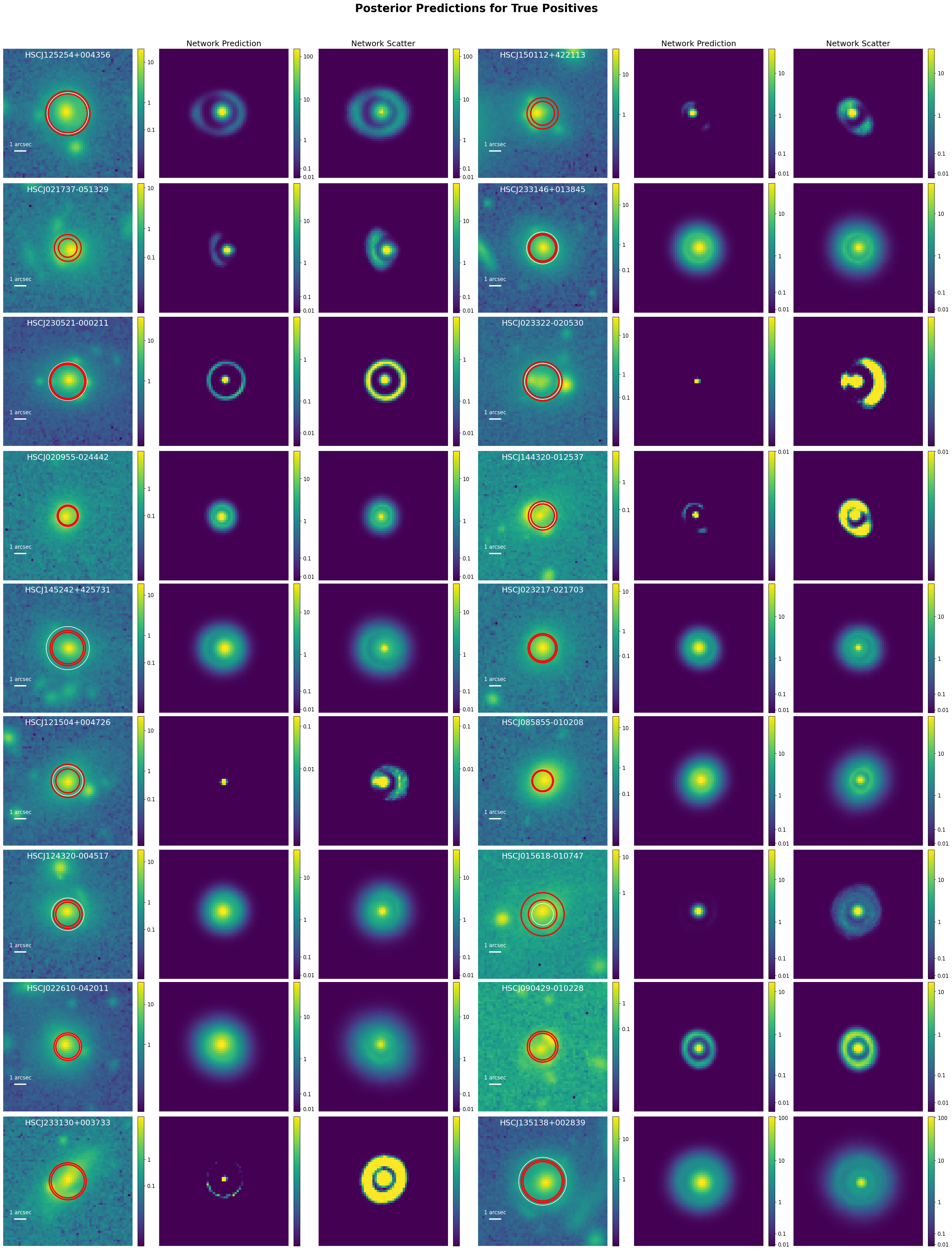}
\caption{\texttt{paltas} posterior images for lens candidates from HSC, based on random draws from the lens parameter posteriors. Left - Right: Original image, per-pixel median prediction, per-pixel scatter in image prediction. The red annuli show the \NNfour{} $\theta_{\rm{E}}$ prediction with width $2\sigma$ and the white circles show Einstein radii values drawn from the literature for comparison.}
\label{Posterior_Predictions_TPs}
\end{figure*}

\begin{figure*}
 \centering
  \centering
  \includegraphics[width=0.95\textwidth]{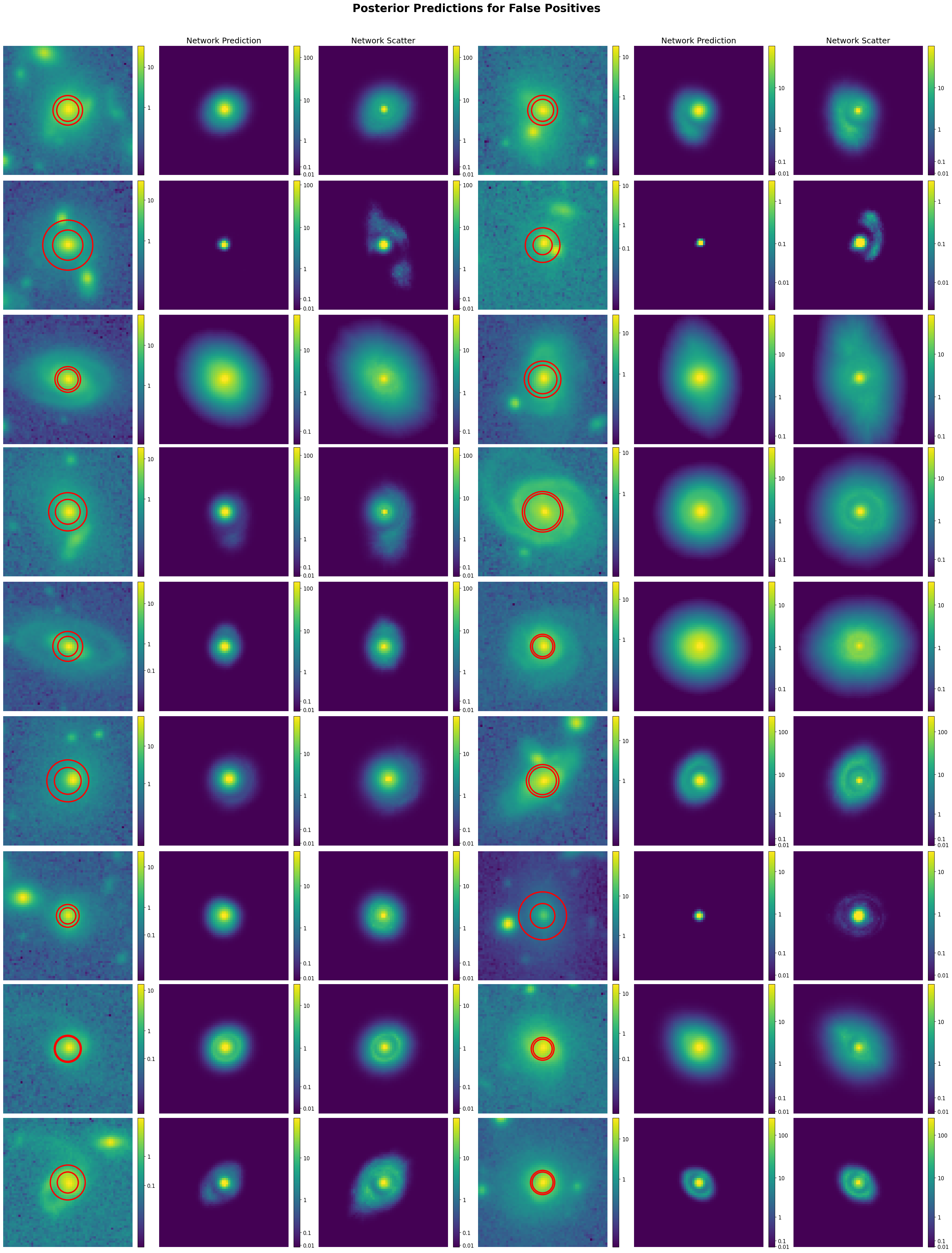}
\caption{\texttt{paltas} posterior images for false positives (non-lenses) from HSC, based on random draws from the lens parameter posteriors. Left - Right: Original image, per-pixel median prediction, per-pixel scatter in image prediction. The red annuli show the \NNfour{} Einstein radius prediction with width $2\sigma$.}
\label{Posterior_Predictions_FPs}
\end{figure*}

\begin{figure*}
 \centering
  \centering
  \includegraphics[width=0.95\textwidth]{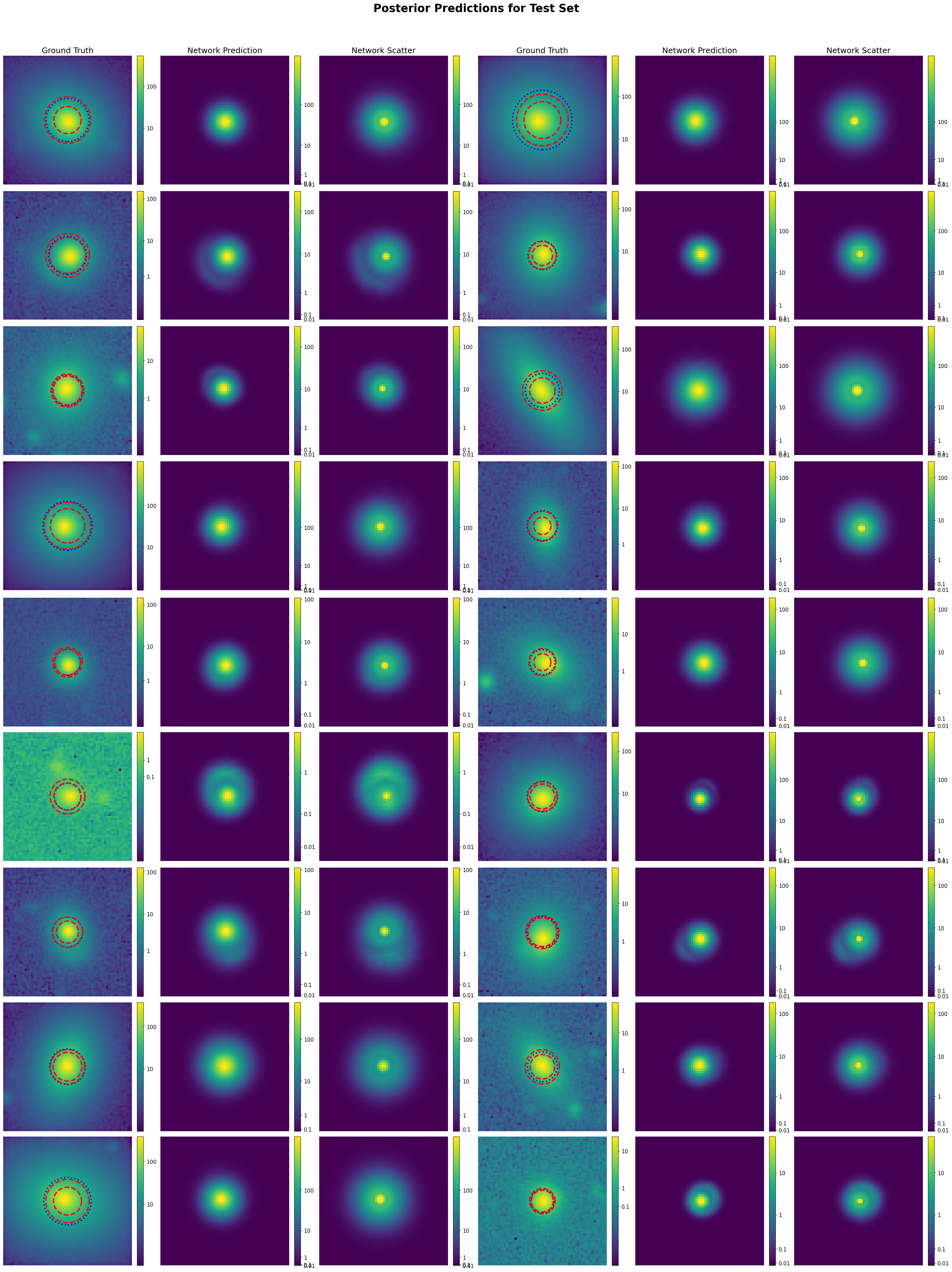}
\caption{\texttt{paltas} posterior images for the simulated LSST test set based on random draws from the lens parameter posteriors. Left - Right: Original image, per-pixel median prediction, per-pixel scatter in image prediction. In the cutout image, the network predicted Einstein radius is plotted (red) with a $2\sigma$ annulus along with the true value (blue).}
\label{Posterior_Predictions_Test}
\end{figure*}

Figures \ref{Posterior_Predictions_TPs}, \ref{Posterior_Predictions_FPs} and \ref{Posterior_Predictions_Test} show a sample of generated posterior images for the true A-grade HSC lenses, the false positives and the simulated LSST test set respectively. The Einstein radii from the literature (white) or catalog ground truth (blue) are shown where relevant, in comparison to the predictions from the network (red annuli). There is large pixel-by-pixel scatter in these posterior images but the median predicted lensed image typically encompasses the lensed arc. Lens systems with clear arcs (e.g., top left Fig. \ref{Posterior_Predictions_TPs}) show good agreement with the original HSC cutout. In the case of the false positives, the per-pixel scatter reveals the network's attempts to recreate the HSC image (such as in row-2,column-1 of Figure \ref{Posterior_Predictions_FPs}) in which neighboring objects in the field are confused for the lensed source. This produces a systematic difference in behavior of the network to true lens systems compared to false positives, as demonstrated in Figures \ref{HSC_FP_Predictions} and \ref{FP_Behaviour_All}.

\bsp	
\label{lastpage}
\end{document}